\documentclass[12pt,a4paper]{article}

\usepackage{amsmath}
\usepackage{amssymb}
\usepackage{epsfig}
\usepackage{graphicx}
\usepackage{cite}

\newcommand{\capdef}{}
\newcommand{\mycaption}[2][\capdef]{\renewcommand{\capdef}{#2}%
       \caption[#1]{{\footnotesize #2}}}
\makeatletter
\renewcommand{\fnum@table}{\textbf{\tablename~\thetable}}
\renewcommand{\fnum@figure}{\textbf{\figurename~\thefigure}}
\makeatother
\def\ltap{\ \raisebox{-.4ex}{\rlap{$\sim$}} \raisebox{.4ex}{$<$}\ }
\def\gtap{\ \raisebox{-.4ex}{\rlap{$\sim$}} \raisebox{.4ex}{$>$}\ }

\newcounter{myenumi}

\renewcommand{\themyenumi}{\roman{myenumi}}
{\end{list}}

\newlength{\myem}
\newcounter{mysubequation}[equation]

\makeatletter
\renewcommand{\section}{\@startsection{section}{1}{0em}{-\baselineskip}%
{\baselineskip}{\normalfont\large\bfseries}}
\renewcommand{\subsection}%
{\@startsection{subsection}{2}{0em}{-0.7\baselineskip}%
{0.7\baselineskip}{\normalfont\bfseries}}
\makeatother

\def\be{\begin{equation}}
\def\ee{\end{equation}}
\newcommand{\beq}{\begin{equation}}
\newcommand{\eeq}{\end{equation}}
\newcommand{\ba}{\begin{array}{c}}
\newcommand{\baz}{\begin{array}{cc}}
\newcommand{\bad}{\begin{array}{ccc}}
\newcommand{\bav}{\begin{array}{cccc}}
\newcommand{\ea}{\end{array}}
\newcommand{\bea}{\begin{equation} \begin{array}{c}}
\newcommand{\eea}{ \end{array} \end{equation}}

\newcommand{\deltaatm}{\mbox{$\Delta m^2_{31}$}}
\newcommand{\deltasol}{\mbox{$ \Delta m^2_{21}$}}
\def\ltap{\ \raisebox{-.4ex}{\rlap{$\sim$}} \raisebox{.4ex}{$<$}\ }
\def\gtap{\ \raisebox{-.4ex}{\rlap{$\sim$}} \raisebox{.4ex}{$>$}\ }
\def\ie{\hbox{\it i.e.}{}}
\def\eg{\hbox{\it e.g.}{}}
\def\etc{\hbox{\it etc}{}}

\newcommand{\dma}{\Delta m_{31}^2}

\newcommand{\Smuhigh}{$S_\mu^\mathrm{high}$}
\newcommand{\Smu}{$S_\mu$}
\newcommand{\Se}{$S_e$}

\begin{document}


\renewcommand{\thefootnote}{\alph{footnote}}

\begin{flushright}
SISSA 94/2005/EP\\
\end{flushright}

\vspace*{1cm}

\renewcommand{\thefootnote}{\fnsymbol{footnote}}
\setcounter{footnote}{-1}

{\begin{center} 
{\Large\textbf{Determining the Neutrino Mass Hierarchy
with Atmospheric Neutrinos}}
\end{center}}

\renewcommand{\thefootnote}{\it\alph{footnote}}

\vspace*{.8cm}

{\begin{center} {{\bf
                S.\ T.\ Petcov\footnote[1]{Also at: Institute of 
                Nuclear Research and Nuclear Energy,
                Bulgarian Academy of Sciences, 1784 Sofia, Bulgaria.
                \makebox[1.cm]{Email:}
                {\sf petcov@he.sissa.it}}
and
                T.\ Schwetz\footnote[2]{\makebox[1.cm]{Email:}
                \sf schwetz@sissa.it}
                }}

\end{center}}
{\it
\begin{center}
       Scuola Internazionale Superiore di Studi Avanzati\\
       Via Beirut 2--4, I--34014 Trieste, Italy
\end{center}}

\vspace*{0.5cm}

\begin{abstract}

The possibility to determine 
the type of neutrino mass hierarchy by studying 
atmospheric neutrino oscillations
with a detector capable to distinguish between 
neutrino and antineutrino events, such as
magnetized iron calorimeters, is considered.
We discuss how the ability to distinguish 
between the neutrino mass spectrum with 
normal and inverted hierarchy depends 
on detector characteristics
like neutrino energy and direction resolutions 
or charge miss-identification,
and on the systematical uncertainties 
related to the atmospheric neutrino fluxes. 
We show also how the neutrino 
mass hierarchy determination depends on the true 
values of $\theta_{13}$ and $\theta_{23}$, 
as well as on the type of the true hierarchy.
We find that for $\mu$-like events, an accurate 
reconstruction of the energy and direction of the 
neutrino greatly improves the sensitivity to the 
type of neutrino mass spectrum.
For $\sin^22\theta_{13} \cong 0.1$ and a
precision of 5\% in the reconstruction of 
the neutrino energy and $5^\circ$ in the neutrino direction, 
the type of neutrino mass hierarchy
can be identified at the 2$\sigma$~C.L.\ with 
approximately 200 events. For resolutions 
of 15\% for the neutrino energy and $15^\circ$ for the neutrino
direction roughly one order of magnitude larger event
numbers are required. 
For a detector capable to distinguish between $\nu_e$ and $\bar\nu_e$
induced events the requirements on energy and direction resolutions
are, in general, less demanding than for a detector with muon charge
identification.
\end{abstract}

\newpage

\renewcommand{\thefootnote}{\arabic{footnote}}
\setcounter{footnote}{0}


\section{Introduction}
\indent 

There has been a remarkable  
progress in the studies of neutrino
oscillations in the last several years.
The experiments with solar, atmospheric and reactor
neutrinos~\cite{sol,SKsolaratm,Ashie:2005ik,SNO123,KamLAND}
have provided compelling evidences for 
existence of neutrino oscillations 
driven by nonzero neutrino masses and neutrino mixing.
Evidences for oscillations of neutrinos were
obtained also in the first long-baseline
accelerator neutrino experiment K2K~\cite{K2K}.
The interpretation of the solar and
atmospheric neutrino, and of KamLAND 
data in terms of 
neutrino oscillations requires
the presence of 3-neutrino mixing
in the weak charged lepton current 
(see, \eg,~\cite{STPNu04}): 
\begin{equation}
\nu_{l \mathrm{L}}  = \sum_{j=1}^{3} U_{l j} \, \nu_{j \mathrm{L}} 
\,,\qquad l  = e,\mu,\tau \,.
\label{3numix}
\end{equation}
%
Here, $\nu_{lL}$ are the three left-handed flavor 
neutrino fields, $\nu_{j \mathrm{L}}$ is the 
left-handed field of the 
neutrino $\nu_j$ with the mass $m_j$
and $U$ is the Pontecorvo-Maki-Nakagawa-Sakata (PMNS)
neutrino mixing matrix~\cite{BPont57},
\begin{equation}
U = \left(\begin{array}{ccc}
U_{e1}& U_{e2} & U_{e3} \\
U_{\mu 1} & U_{\mu 2} & U_{\mu 3} \\
U_{\tau 1} & U_{\tau 2} & U_{\tau 3} 
\end{array} \right)
= \left(\begin{array}{ccc} 
c_{12}c_{13} & s_{12}c_{13} & s_{13}e^{-i\delta}\\
 - s_{12}c_{23} - c_{12}s_{23}s_{13}e^{i\delta} & 
c_{12}c_{23} - s_{12}s_{23}s_{13}e^{i\delta} & s_{23}c_{13}\\
s_{12}s_{23} - c_{12}c_{23}s_{13}e^{i\delta} 
& -c_{12}s_{23} - s_{12}c_{23}s_{13}e^{i\delta}
& c_{23}c_{13}\\ 
\end{array} \right)
\label{Umix}
\end{equation}
%
\noindent where we have used a standard parametrization 
of $U$ with the usual notations, $s_{ij} \equiv \sin \theta_{ij}$,
$c_{ij} \equiv \cos \theta_{ij}$, 
and $\delta$ is the Dirac CP-violation phase
\footnote{We have not 
written explicitly the two possible Majorana 
CP-violation phases~\cite{BHP80,Doi81}
which do not enter into the expressions for the oscillation 
probabilities of interest~\cite{BHP80,Lang87}.}.
If one identifies $\Delta m^2_{21} > 0$ and $\Delta m^2_{31}$
(or  $\Delta m^2_{32}$)  
with the neutrino mass squared differences
which drive the solar and atmospheric 
neutrino oscillations, 
$\theta_{12}$ and $\theta_{23}$
represent the solar and atmospheric 
neutrino mixing angles, 
while $\theta_{13}$ is the angle 
limited by the data from
the CHOOZ and Palo Verde experiments 
\cite{CHOOZPV}. It follows from the data that 
$|\Delta m^2_{31(2)}| \gg \deltasol$.
In this case, obviously, $\Delta m^2_{31} \cong \Delta m^2_{32}$.

  The existing solar and reactor 
neutrino oscillation data allow us to determine 
$\Delta m^2_{21}$ and $\sin^2\theta_{12}$
with a relatively good precision and to obtain
rather stringent limits on $\sin^2\theta_{13}$ (see,
\eg,~\cite{SKsolaratm,Ashie:2005ik,SNO123,KamLAND,BCGPRKL2,3nuGlobal,TSchwU05}).  
The best fit values and the 99.73\% C.L. allowed
ranges of $\Delta m^2_{21}$ and $\sin^2\theta_{12}$
are given by
\footnote{The data imply, in particular,
  that maximal solar neutrino mixing is ruled out at $\sim 6\sigma$.
  At 95\% C.L.\ one finds $\cos 2\theta_\odot \geq
  0.26$~\cite{BCGPRKL2}, which has important
  implications~\cite{PPSNO2bb}.}:
\beq
\label{bfvsol}
\ba
\deltasol = 8.0\times 10^{-5}~{\rm eV^2},~~
\sin^2\theta_{12} = 0.31~, \\[0.25cm]
\deltasol = (7.1 - 9.0) \times 10^{-5}~{\rm eV^2},~~
\sin^2 \theta_{12} = (0.24 - 0.40)~,
\ea
\eeq
It follows from a combined 3-$\nu$ oscillation
analysis of the solar neutrino, 
KamLAND and CHOOZ data that~\cite{BCGPRKL2,TSchwU05}
%
\begin{equation}
\sin^2\theta_{13} < 0.044~~~\mbox{at}~99.73\%~{\rm C.L.}\;
\label{th13}
\end{equation}
%
Existing atmospheric neutrino data
are essentially insensitive to $\theta_{13}$ 
obeying the CHOOZ limit.
The Super-Kamiokande and K2K 
experimental results are best 
described in terms of dominant 
$\nu_{\mu} \rightarrow \nu_{\tau}$ 
($\bar{\nu}_{\mu} \rightarrow \bar{\nu}_{\tau}$)
vacuum oscillations. The best fit values 
and the 99.73\% C.L.\ allowed ranges of the 
atmospheric neutrino oscillation parameters read~\cite{TSchwU05}:
%
\beq 
\label{eq:range}
\ba
|\deltaatm| = 2.2\times 10^{-3}~{\rm eV^2},~ 
~~~\sin^2\theta_{23} = 0.50~,\\  [0.3cm]
|\deltaatm| =(1.4 - 3.3)\times 10^{-3}{\rm eV^2},~~~~
\sin^2\theta_{23} = (0.34 - 0.68)\;.
\ea
\eeq
%
The sign of $\deltaatm$ and of 
$\cos2\theta_{23}$ if
$\sin^22\theta_{23} < 1.0$
cannot be determined using
the existing data. 
The two possibilities,
$\deltaatm > 0$ or $\deltaatm < 0$,
correspond to two different
types of neutrino mass spectrum:
with normal hierarchy (NH),
$m_1 < m_2 < m_3$, and 
with inverted hierarchy (IH),
$m_3 < m_1 < m_2$.  
The fact that the sign of 
$\cos2\theta_{23}$ 
is not determined when
$\sin^22\theta_{23} < 1.0$ implies
that if, \eg, 
$\sin^22\theta_{23} = 0.92$,
two values of  $\sin^2\theta_{23}$ are possible,
$\sin^2\theta_{23} \cong 0.64~{\rm or}~ 0.36$.

  The neutrino oscillation parameters
$\deltasol$, $\sin^2\theta_{12}$,
$|\deltaatm|$ and $\sin^2\theta_{23}$
are determined by the existing
data with a 3$\sigma$ error
of approximately 12\%, 27\%,
50\% and 34\%, respectively.
These parameters can (and very
likely will) be measured with much
higher accuracy in the future (see,
$\eg$,~\cite{STPNu04,TMU04,TSchwU05}).
The MINOS experiment \cite{MINOS}, 
which collects data at present,
will reduce considerably 
(approximately by a factor of 3) the 
current uncertainty in the value of
$|\Delta m^2_{31}|$.
The highest precision in the determination
of $|\Delta m^2_{31}|$
is expected to be achieved 
from the studies
of $\nu_{\mu}$-oscillations in
the T2K experiment 
with the Super-Kamiokande detector~\cite{T2K}:
if the true $|\Delta m^2_{31}| =
2\times 10^{-3}$~eV$^2$ 
(and true $\sin^2\theta_{23} = 0.5$), 
the 3$\sigma$ uncertainty
in $|\dma|$ is estimated to be 
reduced in this experiment to  
$\sim 12\%$~\cite{TMU04}. The error 
diminishes with increasing of
$|\Delta m^2_{31}|$.
The T2K experiment 
will measure also $\sin^2\theta_{23}$
with an error of approximately 14\% at 2$\sigma$
\cite{TSchwU05}. However, this experiment 
would not be able to resolve the 
$\theta_{23}$--$(\pi/2 - \theta_{23})$
ambiguity if $\sin^22\theta_{23}<1$.
In what regards the CHOOZ angle
$\theta_{13}$, there are several 
proposals for reactor $\bar{\nu}_e$ 
experiments with baseline $L\sim$~(1--2)~km 
\cite{Reacth13,Whiteth13}, which could improve 
the current limit by a factor of (5--10).
The reactor $\theta_{13}$ 
experiments can compete in sensitivity with
accelerator experiments 
T2K~\cite{T2K}, NO$\nu$A~\cite{NOvA} 
(see, \eg,~\cite{TMU04}),
and can be done on a relatively short 
time scale. 
The most precise measurement 
of  $\Delta m^2_{21}$ could be achieved 
\cite{SKGdCP04} using Super-Kamiokande 
doped with 0.1\% of gadolinium (SK-Gd) 
for detection of reactor $\bar{\nu}_e$ 
\cite{SKGdBV04}: getting the same flux of 
reactor $\bar{\nu}_e$ as KamLAND,
the SK-Gd detector will have 
approximately a 43 times bigger 
reactor $\bar{\nu}_e$ event rate than KamLAND.
After 3 years of data-taking with SK-Gd, 
$\Delta m^2_{21}$ could be determined with  
an error of 3.5\% at 3$\sigma$~\cite{SKGdCP04}. 
A dedicated reactor $\bar{\nu}_e$ experiment with a 
baseline $L\sim 60$ km, tuned to the minimum of the
$\bar{\nu}_e$ survival probability, 
could provide the most precise 
determination of $\sin^2\theta_{12}$~\cite{TH12,BCGPTH1204}:
with statistics of $\sim 60$ GW~ktyr 
and a systematic error 
of 2\% (5\%), $\sin^2\theta_{12}$  
could be measured with 
an error of 6\% (9\%) at 3$\sigma$~\cite{BCGPTH1204}. 

    Getting more precise 
information about the value 
of the mixing angle $\theta_{13}$,
and measuring the values of $\Delta m^2_{21}$, $|\deltaatm|$,
$\sin^2\theta_{12}$ and $\sin^2\theta_{23}$ 
with a higher precision, 
is crucial for the future progress 
in the studies of neutrino mixing 
(see, \eg,~\cite{STPNu04} 
and the references quoted therein).
Here we focus in particular on the question whether the neutrino mass
spectrum is with normal or inverted hierarchy, \ie, the sign of
$\deltaatm$, which is of fundamental importance for the understanding
of the origin of neutrino masses and mixing. Information on
sign($\deltaatm$) can be obtained through matter effects in
long-baseline experiments~\cite{NOvA,AMMS99,LBL}, or maybe from
supernova neutrinos~\cite{supernova}. If neutrinos with definite mass
are Majorana particles, information about sgn($\deltaatm$) could be
obtained also by measuring the effective Majorana mass in
neutrinoless double $\beta$-decay experiments~\cite{PPSNO2bb,BPP1}.

   In the present article we
investigate in detail the possibility
to obtain information 
on the sign of $\deltaatm$ 
using the data on atmospheric neutrinos, 
which can be obtained in experiments
with detectors able to measure the charge of the muon
produced in the charged current reaction
by atmospheric $\nu_{\mu}$ or $\bar{\nu}_{\mu}$.
Thus, in contrast to water-\v{C}erenkov detectors,
in these experiments
it will be possible to distinguish between 
the $\nu_{\mu}$ and $\bar{\nu}_{\mu}$ induced events.
Among the operating detectors, MINOS
has muon charge identification
capabilities for multi-GeV muons~\cite{MINOS}.
The MINOS experiment is currently collecting 
atmospheric neutrino data. 
The detector has a relatively small mass,
but after 5 years of data-taking 
it is expected to collect about 440 
atmospheric $\nu_{\mu}$ 
and about 260 atmospheric $\bar{\nu}_{\mu}$ 
multi-GeV events. The ATLAS and CMS experiments 
under preparation at the LHC accelerator at CERN
will have efficient muon charge identification 
for multi-GeV muons and, in principle, 
can also be used for studies of atmospheric 
neutrino oscillations in the periods 
when the LHC accelerator will not 
be operational \footnote{We thank F.~Vannucci 
for pointing out this possibility to us.}.
There are also plans to build a 30--50 kton
magnetized iron tracking calorimeter 
detector in India within the India-based 
Neutrino Observatory (INO) 
project~\cite{INO}. The INO detector will 
be based on MONOLITH design~\cite{MONOLITH}. 
The primary goal is to study the oscillations of  
atmospheric $\nu_{\mu}$ and $\bar{\nu}_{\mu}$.
This detector is planned to have
efficient muon charge identification,
high muon energy resolution ($\sim 5\% $) and 
a muon energy threshold of about 2 GeV.
It will accumulate sufficiently high statistics
of atmospheric $\nu_{\mu}$ and $\bar{\nu}_{\mu}$
induced events in several years, which would permit 
to perform a high precision study 
of the $\nu_{\mu}$ and $\bar{\nu}_{\mu}$
disappearance due to oscillations and 
to search for effects of the sub-dominant
$\nu_{\mu} \rightarrow \nu_e$ 
($\nu_{e} \rightarrow \nu_{\mu}$) and
$\bar{\nu}_{\mu} \rightarrow \bar{\nu}_e$
($\bar{\nu}_{e} \rightarrow \bar{\nu}_{\mu}$)
transitions. We consider a detector which 
permits to reconstruct the initial 
$\nu_{\mu}$ ($\bar{\nu}_{\mu}$) energy and
direction with a relatively good precision.
The magnetized iron detectors can have such 
capabilities. We present results also 
for detectors permitting to study 
the oscillations of the atmospheric multi-GeV
$\nu_{e}$ and $\bar{\nu}_{e}$ under similar 
conditions.
 
  The possibility to obtain information about 
the type of neutrino mass spectrum, \ie, 
about the ${\rm sgn}(\deltaatm)$, by investigating 
atmospheric neutrino oscillations 
is based on the prediction that 
in the case of 3-neutrino oscillations,
the Earth matter affects in a different way the 
transitions of neutrinos and antineutrinos~\cite{LW78,BPPW80,MS85}.
For $\deltaatm > 0$, the $\nu_{\mu} \rightarrow \nu_{e}$
($\nu_{e} \rightarrow \nu_{\mu}$) 
transitions of the multi-GeV neutrinos are 
amplified, while the transitions of antineutrinos 
$\bar{\nu}_{\mu} \rightarrow \bar{\nu}_{e}$
($\bar{\nu}_{e} \rightarrow \bar{\nu}_{\mu}$)
are suppressed. If $\deltaatm < 0$, the
indicated transitions of neutrinos are suppressed 
and those of antineutrinos are enhanced. 
The magnitude of the Earth matter effects 
in the oscillations of multi-GeV atmospheric
$\nu_{\mu}$ ($\bar{\nu}_{\mu}$) and
$\nu_{e}$ ($\bar{\nu}_{e}$) depends 
critically on the value of $\sin^2\theta_{13}$:
the effects 
can be substantial for $\sin^2\theta_{13}\gtap 0.01$;
for $\sin^2\theta_{13} \ll 0.01$ they are 
exceedingly small. The difference between the oscillations
of neutrinos and antineutrinos can be relatively large 
and observable in the samples of 
$\mu^{\pm}$ events with $E_{\mu} \sim (2 - 10)$~GeV, 
from atmospheric 
$\nu$'s with a large path length
in the Earth, crossing deeply the 
Earth mantle~\cite{AMMS99} or the mantle and the 
core~\cite{SP3198,SPNu98,106,107,core}.
These are the neutrinos for which
$\cos\theta_n \gtap (0.3 - 0.4)$, where
$\theta_n$ is the Nadir angle
characterizing the neutrino trajectory in the Earth.

Let us note that three-flavour effects in atmospheric neutrino
oscillations including the Earth matter effects have been widely
studied in the literature; for an incomplete list 
see~\cite{core,JBSP203,Palomares-Ruiz:2004tk,%
Fogli:1996nn,Akhmedov,bernabeu,Peres:2003wd,%
ConchaMal02,HyperK,lisi,%
Huber:2005ep,Gandhi:2005wa,Taba02,%
Indumathi:2004kd,Gandhi:2004bj,Choubey:2005zy}. 
A rather detailed analysis for the MONOLITH 
detector has been performed in~\cite{Taba02}, 
and magnetized iron detectors in general have been
considered previously
in~\cite{Palomares-Ruiz:2004tk,Indumathi:2004kd,%
Gandhi:2004bj,Choubey:2005zy}.
A large number of studies have been done for water-\v{C}erenkov
detectors~\cite{JBSP203,ConchaMal02,HyperK,lisi,Huber:2005ep,Gandhi:2005wa}.
Such detectors have a more limited sensitivity to the type of
neutrino mass hierarchy, because only the sum of neutrino and
antineutrino induced events can be observed. Since large matter
effects occur for neutrino if $\deltaatm > 0$ and for antineutrinos if
$\deltaatm < 0$, summing the corresponding events significantly
diminishes the effect of changing the sign of $\deltaatm$. The
remaining sensitivity (see \eg,~\cite{JBSP203,Huber:2005ep}) is due to
the facts that neutrino and antineutrino cross sections differ roughly
by a factor of 2, and that a huge number of events is available in
mega ton scale experiments~\cite{HyperK}.
Obviously, such a ``dilution'' of the magnitude of the matter effects
does not take place in the samples of $\mu^-$ and
$\mu^+$ events, which can be collected in the experiments with muon
charge identification.

\section{Three-Neutrino Oscillations of Atmospheric Neutrinos}
\indent

In the present Section we will review briefly
the formalism and the physics of the 3-neutrino 
oscillations of the multi-GeV atmospheric neutrinos 
with energy $E \sim (1 - 10)$ GeV in the Earth 
(see, \eg,~\cite{JBSP203} and the references 
quoted therein for further details).  
The probabilities of the flavour neutrino oscillations
in matter of interest, $\nu_{\alpha} \to \nu_{\beta}$,
$\bar{\nu}_{\alpha} \to \bar{\nu}_{\beta}$,  
$\alpha=e,\mu$, $\beta=e,\mu,\tau$,
can be written for an arbitrary matter density profile as
\begin{equation}\label{eq:prob}
P^p_{\alpha\beta} = |A^p_{\alpha\beta}|^2 \,, 
\end{equation}
%
where $A^p_{\alpha\beta}$ is 
the probability amplitude which is
obtained by solving a Schr{\"o}dinger-like 
system of coupled evolution equations and
$p = \nu~(\bar\nu)$ for neutrinos (antineutrinos). 
We use the standard parameterization of 
the PMNS mixing matrix \cite{3nuKP88},
$U=O_{23} U_\delta O_{13} O_{12}$, 
where $O_{ij}$ is the orthogonal matrix of 
rotations in the $ij$-plane which depends on the
mixing angle $\theta_{ij}$, and $U_\delta=\mathrm{diag} (1,1,{\rm
e}^{{\rm i}\delta})$, 
$\delta$ being the Dirac CP-violating phase. 
Exploiting the fact that the 
matrix $O_{23}U_\delta$ commutes with the
matrix containing the matter potential, 
it is easy to show that \cite{3nuSP88}
without loss of generality $A^p_{\alpha\beta}$ 
can be written as
\begin{equation}
A^p = O_{23}U_\delta A'^p U^\dagger_\delta O_{23}^T \,,
\label{eq:AA'}
\end{equation}
%
where $A'^p$ does not depend on $\theta_{23}$ and $\delta$. We adopt
further the approximation of setting the neutrino mass-squared
difference $\Delta m^2_{21}$, responsible for the solar neutrino
oscillations, to zero
\footnote{This can be justified by the fact that
we are interested in oscillations of
multi-GeV (atmospheric) neutrinos 
in the Earth and that the existing data 
implies $\deltasol/|\deltaatm| \cong 0.04 \ll 1$.  
We have checked by explicit calculations that 
for the energy range of interest, $E_\nu \gtap 1$~GeV, 
our results are not affected by setting $\Delta m^2_{21} = 0$
if $\Delta m^2_{21}$ has values in the interval
given in Eq.~(\ref{bfvsol}).}. 
In this approximation 
the probabilities of interest are independent of the 
mixing angle $\theta_{12}$ and the Dirac
CP-violating phase $\delta$. Moreover, the calculation 
of $A'^p$ is reduced effectively to the calculation of
probability amplitudes in the case of two neutrino
mixing~\cite{3nuSP88}. 
For neutrinos with energy $E$ and crossing the Earth along a
trajectory characterized by a Nadir angle $\theta_{n}$, the probability
of the 2-neutrino $\nu_{e} \rightarrow \nu'$ ($\bar{\nu}_{e}
\rightarrow \bar{\nu}'$) transitions is given by
\begin{equation}
P^{2p} \equiv P^{2p}(\deltaatm, \theta_{13};E,\theta_{n})
\equiv |A'^p_{13}|^2 = |A'^p_{31}|^2 = 
1 - |A'^p_{11}|^2 = 1 - |A'^p_{33}|^2 \,,
\end{equation}
where $p=\nu~(\bar{\nu})$, and $\nu' = s_{23}\nu_{\mu} + c_{23}
\nu_{\tau}$~\cite{3nuSP88}. The three-neutrino oscillation
probabilities are obtained as
%
\begin{eqnarray}
P^{3p}_{e e} &\cong& |A'^p_{11}|^2 = 1 - P^{2p}\,,
  \label{P3ee} \\
P^{3p}_{\mu e} = P^{3p}_{e \mu} 
  &\cong& s_{23}^2 \, |A'^p_{13}|^2 = s_{23}^2 \, P^{2p}\,,
  \label{P3emu} \\
P^{3p}_{e \tau } &\cong& c_{23}^2 \, |A'^p_{13}|^2
  = c_{23}^2 \, P^{2p}\,, \label{P3etau} \\
P^{3p}_{\mu\mu} & \cong &
  c^4_{23} + s^4_{23} \, |A'^p_{33}|^2 + 
  2c^2_{23}s^2_{23} \, \mathrm{Re}(A'^p_{33}) \\ 
  & = & 1 - s_{23}^4 \, P^{2p} 
  - 2c^2_{23}s^2_{23} \, \left [ 1 -
  \mathrm{Re}( e^{-i\kappa_{p}} A^{2p}_{\nu'\nu'}) \right ]\,,
  \label{P3mumu} \\
P^{3p}_{\mu \tau } &=& 
  1 - P^{3p}_{\mu \mu} - P^{3p}_{\mu e} \,. \label{P3mutau}
\end{eqnarray}
%
In Eq.~(\ref{P3mumu}) we used $e^{-i\kappa_{p}}A^{2p}_{\nu'\nu'}
\equiv A'^p_{33}$, where $\kappa_{p}$ and $A^{2p}_{\nu' \nu'}$ are a
known phase and 2-neutrino transition probability
amplitude~\cite{3nuSP88,SP3198,SPNu98}. 

 The fluxes of atmospheric  $\nu_{e,\mu}$ and
$\bar{\nu}_{e,\mu}$
of energy $E$, which reach the detector after
crossing the Earth along a given trajectory  
specified by the value of $\theta_{n}$
are given by the following expressions 
in the case of the 3-neutrino oscillations  
under discussion~\cite{SPNu98}:
\begin{eqnarray}
\bar{\Phi}^{p}_{e}(E,\theta_{n}) &\cong& 
  \Phi^p_e \, \left[ 1 + 
  (s^2_{23}r^p - 1) \, P^{2p}\right] \,,
  \label{Phie}\\
\bar{\Phi}^{p}_{\mu}(E,\theta_{n}) &\cong& 
  \Phi^{p}_{\mu} \, \left\{ 1 +
  s^4_{23}\, [(s^2_{23}~r^p)^{-1} - 1] \, P^{2p}
  - 2c^2_{23}s^2_{23} \,
  \left[ 1 - \mathrm{Re}(A'^p_{33})\right]
  \right\} \,, \label{Phimu}
\end{eqnarray}
%
where 
$\Phi^{p}_{e(\mu)} = \Phi^{p}_{e(\mu)}(E,\theta_{n})$
are the $\nu_{e(\mu)}$ ($p=\nu$) and
$\bar{\nu}_{e(\mu)}$ ($p=\bar{\nu}$) 
fluxes in the absence of neutrino 
oscillations, and
%
\begin{equation}
r^p \equiv r^p(E,\theta_{n}) \equiv
\frac{\Phi^{p}_{\mu}(E,\theta_n)} 
{\Phi^{p}_{e}(E,\theta_n)}\,.
\label{r}
\end{equation}

The interpretation of the 
SK atmospheric neutrino data in terms of 
dominant $\nu_{\mu} \rightarrow \nu_{\tau}$
and $\bar{\nu}_{\mu} \rightarrow \bar{\nu}_{\tau}$
oscillations requires the parameter 
$s^2_{23}$ to lie approximately in the interval
$(0.30 - 0.70)$, with 0.5 being the statistically 
preferred value. For the predicted
ratio $r^p(E,\theta_{n})$ of the fluxes 
of atmospheric $\nu_{\mu}$ ($\bar{\nu}_{\mu}$) and $\nu_e$
($\bar{\nu}_e$) with energy 
$E \cong (2 - 10)$ GeV, 
crossing the Earth along trajectories
for which $0.3 \ltap \cos\theta_{n}\leq 1.0$, 
one has \cite{fluxes}:
$r^p(E,\theta_{n}) \cong (2.6 - 4.5)$. 
Correspondingly, for $s^2_{23} = 0.5~(0.64)$ we find 
for the factors multiplying the 2-neutrino transition
probability $P^{2p}$ in Eqs.~(\ref{Phie}) and (\ref{Phimu}): 
$(s^2_{23}~r^p - 1) 
\cong 0.3 - 1.3~(0.66 - 1.9)$, and
$s^4_{23}[1 - (s^2_{23}~r^p)^{-1}] \cong 
0.06 - 0.14~(0.16 - 0.27)$.
This result implies that the 
2-neutrino transition probability $P^{2p}$ 
can have a much larger effect 
on the flux of atmospheric $\nu_e$ ($\bar{\nu}_e$) 
than on the flux of atmospheric 
$\nu_{\mu}$ ($\bar{\nu}_{\mu}$).

  In our further analysis
performed in Sections 5 and 6,  
the matrix $A'$ and correspondingly
the probabilities $P^{2p}$ and the amplitudes
$A'^p_{33} = e^{-i\kappa_{p}}A^{2p}_{\nu'\nu'}$, 
are obtained by solving
the relevant system of evolution equations
(see, \eg,~\cite{3nuSP88})
numerically using a realistic Earth 
density profile provided by the PREM model~\cite{PREM81}.
As is well-known, the Earth density distribution
in the existing Earth models is assumed to be 
spherically symmetric.
There are two major density structures---the core and the mantle, and 
a certain number of substructures (shells or layers)
\footnote{According to the Earth models
\cite{Stacey:1977,PREM81},
the core has a radius $R_c = 3485.7~$km,
the Earth mantle depth is approximately $R_{man} = 2885.3~$km,
and the Earth radius is $R_{\oplus} = 6371~$km.  
The mean values of the matter densities and the electron fraction 
numbers in the mantle and in the core read, respectively: 
$\bar{\rho}_{man} \cong 4.5~{\rm g/cm^3}$, 
$\bar{\rho}_c \cong 11.5~{\rm g/cm^3}$, and~\cite{Art2}
$Y_e^{man} = 0.49$,  $Y_e^{c} = 0.467$. 
}. 
The mean electron number densities
in the mantle and in the core 
read \cite{PREM81}: $\bar{N}_{e}^{man} 
\cong 2.2~N_A\,\mathrm{cm}^{-3}$, 
$\bar{N}_e^c  
\cong 5.4~N_A \, \mathrm{cm}^{-3}$,
$N_A$ being the Avogadro number. 

  Rather simple analytic expressions for
the quantities of interest, 
$P^{2p}$, $\kappa_{p}$ and $A^{2p}_{\nu'\nu'}$
have been derived in 
the two-layer model of the Earth
density distribution (see, \eg,~\cite{3nuKP88}).
In this model the electron number densities
in the mantle and in the
core are assumed to be constant
\footnote{Numerical calculations 
have shown~\cite{SP3198,MMQLSP97} 
that, \eg, the  
2-neutrino oscillation probability 
$P^{2p}$ of interest, calculated within 
the two-layer model of the Earth
with $\bar{N}_{e}^{man}$ and $\bar{N}_{e}^{c}$
for a given neutrino trajectory
determined using the PREM 
(or the Stacey \cite{Stacey:1977}) model,  
reproduces with a rather high precision the 
corresponding probability, 
calculated by solving numerically the 
relevant system of evolution equations
with the more sophisticated Earth density profile
of the PREM (or Stacey) model.}.  
For the atmospheric neutrinos crossing 
only the Earth mantle but not the Earth core
($\theta_{n} \gtap 33.17^\circ$), for instance,
the expressions for 
$P^{2p}$, $\kappa_p$ and  $A^{2p}_{\nu'\nu'}$
in the two-layer model
are given by (see, \eg,~\cite{JBSP203}):
\begin{eqnarray}
P^{2p}
  &=& \sin^2{\left(\frac{\Delta E'^{p}_{m} L}{2}\right)} 
  \sin^2 2\theta'^{p}_{m} \,,\label{p2n} \\
\kappa_{\nu(\bar{\nu})} &\cong& 
  \frac{1}{2} \, \left[ \frac{\deltaatm L}{2 E}
  \raisebox{-.4ex}{\rlap{{\tiny(}$-${\tiny)}}} \raisebox{.5ex}{$\:+$} \:
  \sqrt{2}G_F\bar{N}_e^{man} L - 
  \Delta E'^{\nu(\bar{\nu})}_{m}L \right] \,, \\
A^{2p}_{\nu'\nu'} &=& 1 +
  \left( e^{-i \Delta E'^{p}_{m}L} - 1 \right) 
  \cos^2\theta'^{p}_{m} \,.
\end{eqnarray}
%
\noindent Here $\Delta E'^{p}_{m}$ 
and $\theta'^{p}_{m}$ are 
the difference between the energies of the 
two matter-eigenstate neutrinos 
and the mixing angle (in matter) 
in the mantle, which coincide with
$\deltaatm/(2E)$ and $\theta_{13}$ in vacuum,
and $L$ is the distance the neutrino
travels in the mantle,
$L = 2 R_{\oplus}\cos \theta_n$, where
$R_{\oplus} = 6371~$km is 
the Earth radius. Similar expressions 
have been derived in the case of neutrinos crossing
the Earth core \cite{SP3198,SPNu98}
(see also \cite{JBSP203}).  

For a given neutrino trajectory in the Earth,
the 2-neutrino transition probability
$P^{2p}$ is a rather slowly varying function of 
the neutrino energy in the interval 
of interest $E\sim (1 - 10)$~GeV.  
In contrast, the interference term 
Re$(e^{-i\kappa_{p}} A^{2p}_{\nu'\nu'})$  
in the expression of the $\nu_{\mu}$
($\bar{\nu}_{\mu}$) survival probability
$P^{3p}_{\mu\mu}$ is a relatively fast
oscillating function of $E$. 
As a consequence, the integration over the 
atmospheric neutrino energy in the indicated 
interval does not affect significantly 
the probability $P^{2p}$, while 
it tends to suppress strongly the
effects due to the term 
Re$( e^{-i\kappa_{p}} A^{2p}_{\nu'\nu'})$.
The specific properties of the 
oscillating term 
Re$( e^{-i\kappa_{p}} A^{2p}_{\nu'\nu'})$, 
present in the expression of the $\nu_{\mu}$
($\bar{\nu}_{\mu}$) survival probability
$P^{3p}_{\mu\mu}$, have been discussed
in the case of neutrinos crossing only the Earth
mantle in~\cite{Gandhi:2004bj}. 
A further discussion of the interplay between the term $\propto
P^{2p}$ and that $\propto \mathrm{Re}(A'^p_{33})$ in the $\nu_{\mu}$
($\bar{\nu}_{\mu}$) survival probability is given in Section 4.

For $\deltaatm > 0$, we can have 
$P_{2\nu} \cong 1$ in the case of
atmospheric neutrinos crossing 
only the mantle for neutrino energy 
and path-length given by
$E_{res} \cong 6.6~\deltaatm[10^{-3}~{\rm eV^2}]~\cos2\theta_{13}$
($\bar{N}_e^{man}[N_A \mathrm{cm}^{-3}])^{-1}~{\rm GeV}$ 
and 
$L[10^4\,{\rm km}] \cong 
(0.8~\bar{N}_e^{man}[N_A \mathrm{cm}^{-3}]\tan2\theta_{13})^{-1}$.
Taking $\deltaatm \cong 2.0 \times 10^{-3}~{\rm eV^2}$
and $\sin^2\theta_{13} = 0.05~(0.025)$
one finds that $E_{res} \cong 6.6~{\rm GeV}$,
and that we can have $P_{2\nu} \cong 1$
only if $L \cong 8000~(10000)~{\rm km}$.  

  In the case of atmospheric 
neutrinos crossing the Earth core, 
it is possible to have $P^{2\nu} \cong 1$ 
for $\sin^2\theta_{13} < 0.05$ and $\deltaatm > 0$,
only due to the effect of maximal constructive 
interference between the amplitudes of the 
the $\nu_{e} \rightarrow \nu'_{\tau}$
transitions in the Earth mantle and in the 
Earth core~\cite{SP3198,SPNu98,106,107}
\footnote{The effect differs from the MSW one~\cite{SP3198} and the 
enhancement happens in the case of interest at 
a value of the energy  between the resonance energies 
corresponding to the density in the mantle 
and that of the core.}.
The {\it mantle-core enhancement effect} 
is caused by the existence 
(for a given $\nu$-trajectory
through the Earth core) of 
{\it points of resonance-like 
total neutrino conversion}, 
$P^{2\nu} = 1$, in the corresponding 
space of $\nu$-oscillation 
parameters~\cite{106,107}. 
The points where $P^{2\nu} = 1$
are determined by two conditions~\cite{106,107}.
A rather complete set of values of 
$\deltaatm/E$ and $\sin^22\theta_{13}$
for which both conditions 
hold and $P^{2\nu} = 1$
was found in~\cite{107}. 
The location of these points determines the regions
where $P^{2\nu}$ is large, $P^{2\nu} \gtap 0.5$. 
For $\sin^2\theta_{13} < 0.05$,
there are two sets of values of 
$\deltaatm$ and $\sin^2\theta_{13}$
for which $P^{2\nu} = 1$:
for, \eg, $\theta_n = 0;~13^0;23^0$,
we have $P^{2\nu} = 1$ at  
1) $\sin^22\theta_{13} = 0.034;~0.039;~0.051$, 
and at 2) $\sin^22\theta_{13} = 0.15;~0.17;~0.22$ 
(see Table~2 in~\cite{107}).
For $\deltaatm = 2.0\times 10^{-3}~{\rm eV^2}$, 
$P^{2\nu} = 1$ occurs in the case 1) 
at $E \cong (2.8 - 3.1)~{\rm GeV}$.
The effects of the mantle-core 
enhancement of $P^{2\nu}$ (or $P^{2\bar{\nu}}$ 
if $\deltaatm < 0$) 
increase rapidly with $\sin^22\theta_{13}$
as long as $\sin^22\theta_{13}\ltap 0.06$,
and should exhibit a rather weak dependence on
$\sin^22\theta_{13}$ for
$0.06 \ltap \sin^22\theta_{13} < 0.19$.

In our further analysis amplitudes of interest $A'^p_{13}$,
$A'^p_{33}$, \etc\ are obtained by solving the evolution equation
numerically using a realistic Earth density profile~\cite{PREM81}.  To
simplify the notation we introduce the abbreviation $\Delta m^2 \equiv
\Delta m^2_{31}$.

\section{Description of the Analysis}
\indent

{\bf Detection cross section.}
In the following we assume detection of atmospheric neutrinos of
flavour $\alpha = e,\mu$ by the charged-current interaction
$\nu_\alpha + N \to \alpha + X$, where $N$ is the detector nucleon,
$\alpha$ is the charged lepton, and $X$ contains all the additional
particles in the final state. Since we are interested in energies
large compared to the muon mass the total cross section for this
reaction is equal for $\nu_e$ and $\nu_\mu$, and we denote it by
$\sigma^p(E_\nu)$, where $p = \nu(\bar\nu)$ for neutrinos
(antineutrinos).  We take into account that in realistic detectors
there will be an energy threshold $E_\mathrm{thr}$ for the detection
of the charged leptons (\eg, muons). Therefore, the relevant cross
section is given by
\begin{equation}
\sigma^p(E_\nu) = \int_{E_\mathrm{thr}}^{E_\nu} dE_\alpha 
\frac{d\sigma^p}{dE_\alpha} \approx 
\sigma^p_\mathrm{tot}(E_\nu) \left[ 1 - \frac{E_\mathrm{thr}}{E_\nu}
\left(\frac{1 + a^p\, E_\mathrm{thr}^2/E_\nu^2}{1+a^p}\right)\right] \,,
\end{equation}
where $\sigma^p_\mathrm{tot}$ is the total cross section for
$E_\mathrm{thr} = 0$, and $a^\nu = \bar Q/3Q$, $a^{\bar\nu} = Q/3\bar
Q$ with $Q \approx 0.44$, $\bar Q \approx 0.06$ being the average
fractions of the momentum of the proton carried by quarks and
anti-quarks, respectively. These numbers hold for a iso-scalar target,
and our results do hardly depend on the precise values adopted for $Q$
and $\bar Q$. In our calculations we use a threshold of
$E_\mathrm{thr} = 2$~GeV.
We assume that the detector provides enough information on momentum
and energy of the final states, such that the initial neutrino energy
and neutrino direction can be reconstructed with an accuracy of
$\sigma_E$ and $\sigma_\mathrm{dir}$, respectively. 

{\bf Atmospheric neutrino fluxes.}
For the initial fluxes of atmospheric neutrinos we use the results of
3-dimensional calculations~\cite{Honda:2004yz}. Since we
consider only upward-going events with $0.1 \le \cos\theta_n \le 1$,
where $\theta_n$ is the Nadir angle of the neutrino, one can neglect
the height distribution of the neutrino production in the atmosphere.
Furthermore, we average the fluxes over the azimuthal angle, such that
we finally obtain fluxes $\Phi_\alpha^p(E_\nu, \cos\theta_n)$, with
$\alpha = e,\mu$ and $p = \nu(\bar\nu)$.

{\bf Calculation of the event rates.}
Using the fluxes, cross sections and oscillation probabilities
discussed earlier, the predicted spectrum of $\alpha$-like events as a
function of the neutrino energy $E_\nu$ and the Nadir angle
$\cos\theta_n$ can be calculated by
\begin{equation}\label{eq:spectrum}
S^p_\alpha(E_\nu, \cos\theta_n) \propto 
\left[
\Phi_\alpha^p(E_\nu, \cos\theta_n)
P_{\alpha\alpha}^p(E_\nu, \cos\theta_n)
+
\Phi_\beta^p(E_\nu, \cos\theta_n)
P_{\beta\alpha}^p(E_\nu, \cos\theta_n)
\right]
\sigma^p(E_\nu)
\end{equation}
where $\alpha,\beta = e,\mu$, $\alpha \neq \beta$. To take into
account the uncertainty in the reconstruction of neutrino energy and
direction we fold this spectrum with Gaussian resolution functions
$R_E$ and $R_\mathrm{dir}$:
\begin{equation}\label{eq:resolutions}
\hat S^p_\alpha(E_\nu, \cos\theta_n) =
\int d E'_\nu \, R_E(E_\nu,E'_\nu)
\int d (\cos\theta'_n) \, R_\mathrm{dir}(\cos\theta_n, \cos\theta'_n) \,
S^p_\alpha(E'_\nu, \cos\theta'_n) \,.
\end{equation}
We assume a constant {\it relative} energy resolution, \ie, $\sigma_E
\propto E_\nu$. Furthermore, we adopt the approximation of an angular
resolution $\sigma_\mathrm{dir}$ being independent of the neutrino
energy. In the resolution function $R_\mathrm{dir}$ we convert
$\sigma_\mathrm{dir}$ into an uncertainty on $\cos\theta_n$ by
$\sigma(\cos\theta_n) = \sigma_\mathrm{dir}\sin\theta_n$ with
$\sigma_\mathrm{dir}$ in radiant.

Departing from the spectrum given in
Eq.~(\ref{eq:resolutions}) we calculate event numbers divided into
$N^\mathrm{bin}_E$ bins in $E_\nu$ between $E_\nu^\mathrm{min}$ and
$E_\nu^\mathrm{max}$, and into $N^\mathrm{bin}_\theta$ bins in
$\cos\theta_n$ between $\cos\theta_n^\mathrm{min}$ and
$\cos\theta_n^\mathrm{max}$:
\begin{equation}\label{eq:rates}
N^p_{jk} = \mathcal{C} \int_j dE_\nu \int_k d(\cos\theta_n)
\hat S^p_\alpha(E_\nu, \cos\theta_n) 
\end{equation}
where the integration boundaries are given by the $j$th bin in
neutrino energy and the $k$th bin in $\cos\theta_n$. The normalization
constant $\mathcal{C}$ is determined by fixing the total number of
neutrino and antineutrino events to $N^\mathrm{tot}$ for specified
values of the oscillation parameters $\boldsymbol{x} =(\theta_{13},
\theta_{23}, \Delta m^2)$:
\begin{equation}\label{eq:normalization}
N^\mathrm{tot} = 
\sum_{j=1}^{N^\mathrm{bin}_E} 
\sum_{k=1}^{N^\mathrm{bin}_\theta} 
\sum_{p=\nu,\bar\nu}
N^p_{jk} (\boldsymbol{x})\,.
\end{equation}
In general we choose for the normalization the ``true'' parameter
values $\boldsymbol{x}^\mathrm{true}$, the same which are used to
simulate the ``data'' (see below).  For the calculation of our actual
observables we take into account that in a realistic experiment charge
identification will not be perfect. Assuming that the charge
assignment is correct for a fraction $f_\mathrm{ID}$ of all events we
obtain the actual observables in each bin for neutrinos and
antineutrinos by
\begin{equation}\label{eq:observables}
\begin{array}{c}
R^\nu_{jk} = 
f_\mathrm{ID}\, N^\nu_{jk} + (1 - f_\mathrm{ID})\, N^{\bar\nu}_{jk} \\ 
R^{\bar\nu}_{jk} = 
f_\mathrm{ID}\, N^{\bar\nu}_{jk} + (1 - f_\mathrm{ID})\, N^\nu_{jk} 
\end{array}
\end{equation}
Hence, 100\% correct charge identification corresponds to
$f_\mathrm{ID}=1$, and $f_\mathrm{ID}=0.5$ implies no information on
the charge, since neutrino and antineutrino events are summed with
equal weight. For simplicity we assume here that $f_\mathrm{ID}$ does
neither depend on the energy nor on the direction.\footnote{For a
given experiment this might be only a rough approximation, since in
general the ability to identify the charge of a particle {\it does}
depend on the energy and on the direction relative to the magnetic
field.}

{\bf The $\chi^2$-analysis and systematic errors.}
We investigate the potential to determine the mass hierarchy by
performing a statistical analysis based on a $\chi^2$-function.
Following the standard approach to construct a $\chi$-function for
future experiments we calculate ``data'' by adopting ``true values''
$\boldsymbol{x}^\mathrm{true}$ for the oscillation parameters:
\begin{equation}\label{eq:data}
D^p_{jk} = R^p_{jk}(\boldsymbol{x}^\mathrm{true}) \,.
\end{equation}
In the theoretical
prediction we take into account several sources of systematic errors
by introducing 12 pull variables $\boldsymbol{\xi} =
(\xi_1,\ldots,\xi_{12})$:
\begin{equation}\label{eq:pred}
T^p_{jk}(\boldsymbol{x},\boldsymbol{\xi} ) = 
R^p_{jk}(\boldsymbol{x}) 
\left( 1 + \sum_{l=1}^{12} \xi_l \, \pi^p_{jk,l} \right)\,,
\end{equation}
with appropriately defined ``couplings'' $\pi^p_{jk,l}$.  In the
systematic error treatment we follow closely the description given in
the appendix of~\cite{Gonzalez-Garcia:2004wg}, where also a definition
of the $\pi^p_{jk,l}$ can be found (see also~\cite{Ashie:2005ik}). The
systematic effects included in our analysis are listed in
Tab.~\ref{tab:systematics}. We consider a fully correlated overall
normalization error from various sources like uncertainties in the
atmospheric neutrino fluxes, the cross sections, the fiducial detector
mass, or efficiencies. We adopt a conservative value of 20\%.
Furthermore, we take into account an uncertainty in the
neutrino/antineutrino ratio (including fluxes as well as cross
sections) and an error on the ratio of $e$-like to $\mu$-like fluxes.
In addition to these normalization errors we allow also for
uncertainties in the {\it shape} of the neutrino fluxes by introducing
a linear tilt in $\cos\theta_n$ and $E_\nu$, uncorrelated between the
four fluxes of $\nu_e, \bar\nu_e, \nu_\mu, \bar\nu_\mu$. And finally
we take into account the effect of an uncertainty in the charge
identification fraction $f_\mathrm{ID}$.
We consider the values for the systematics given in
Tab.~\ref{tab:systematics} as conservative estimates. We will use
these representative values to illustrate the impact of such kind of
uncertainties on the mass hierarchy sensitivity by comparing the
$\chi^2$ with and without including them.

\begin{table}
\centering
\begin{tabular}{|c|l|r|}
\hline\hline
index $l$ & systematic effect & value \\
\hline
1 & overall normalization & 20\% \\
2 & $\nu/\bar\nu$ ratio & 5\% \\
3 & $\nu_\mu/\nu_e$ ratio of fluxes & 5\% \\
$4-7$ & $\cos\theta_n$ dependence of fluxes & 5\%\\
$8-11$ & energy dependence of fluxes & 5\%\\
$12$ & uncertainty on $f_\mathrm{ID}$ & 5\%\\
\hline\hline
\end{tabular}
\mycaption{Systematic uncertainties included in the $\chi^2$ analysis.
\label{tab:systematics}}
\end{table} 

Allowing for rather small event numbers per bin we adopt a
$\chi^2$-definition based on Poisson statistics:
\begin{equation}\label{eq:chisq}
\chi^2(\boldsymbol{x}^\mathrm{true};\boldsymbol{x}) = 
\min_{\boldsymbol{\xi}} 
\left[
2 
\sum_{j=1}^{N^\mathrm{bin}_E} 
\sum_{k=1}^{N^\mathrm{bin}_\theta} 
\sum_{p=\nu,\bar\nu}
\left(
T^p_{jk}(\boldsymbol{x},\boldsymbol{\xi} ) - D^p_{jk} + 
D^p_{jk} \ln \frac{D^p_{jk}}{T^p_{jk}(\boldsymbol{x},\boldsymbol{\xi})}
\right)
+
\sum_{l=1}^{12} \xi_l^2
\right]
\end{equation}
Since we are going to consider up to $N^\mathrm{bin}_E \times
N^\mathrm{bin}_\theta = 20 \times 20$ bins, and the total number of
events is expected to be of order of a few hundred, the actual number
of events per bin will be often less than 1. In a real experiment of
that kind an un-binned likelihood analysis should be performed in order
to extract most information from the available data. In the absence of
real data, however, the $\chi^2$ procedure outlined here still
describes the performance of an ``average'' experiment.


\section{Qualitative Discussion}
\indent

In this Section we give a qualitative discussion of the effects
allowing to distinguish the two types of neutrino mass hierarchy under
the condition of relatively precise measurement of the event spectrum.
This considerations will allow to obtain some insight in the numerical
results presented in the subsequent sections.

Substituting Eqs.~(\ref{Phie}) and (\ref{Phimu}) into
Eq.~(\ref{eq:spectrum}), one obtains for the predicted event spectrum
(see, \eg,~\cite{Palomares-Ruiz:2004tk})
\begin{eqnarray}
S^p_e &\propto&
\Phi_e^p \sigma^p
\left[1 + (r^p \sin^2\theta_{23} -1 ) P^p_{2\nu} \right] 
\,, \label{eq:spectrum_e}\\
S^p_\mu &\propto&
\Phi_\mu^p \sigma^p
\left[ 1 
+ \sin^2\theta_{23}
\left(\frac{1}{r^p} - \sin^2\theta_{23} \right) P^p_{2\nu}
- 
\frac{1}{2}\sin^22\theta_{23} 
\left(1 - \mathrm{Re} \, (A'^p_{33})\right)
\right] \,,\label{eq:spectrum_mu}
\end{eqnarray}
where $r^p\equiv r^p(E_\nu,\cos\theta_n)$ is defined in Eq.~(\ref{r}) 
and we have suppressed the dependence on $E_\nu$ and $\cos\theta_n$.

\begin{figure}[t]
\centering 
\includegraphics[width=0.95\textwidth]{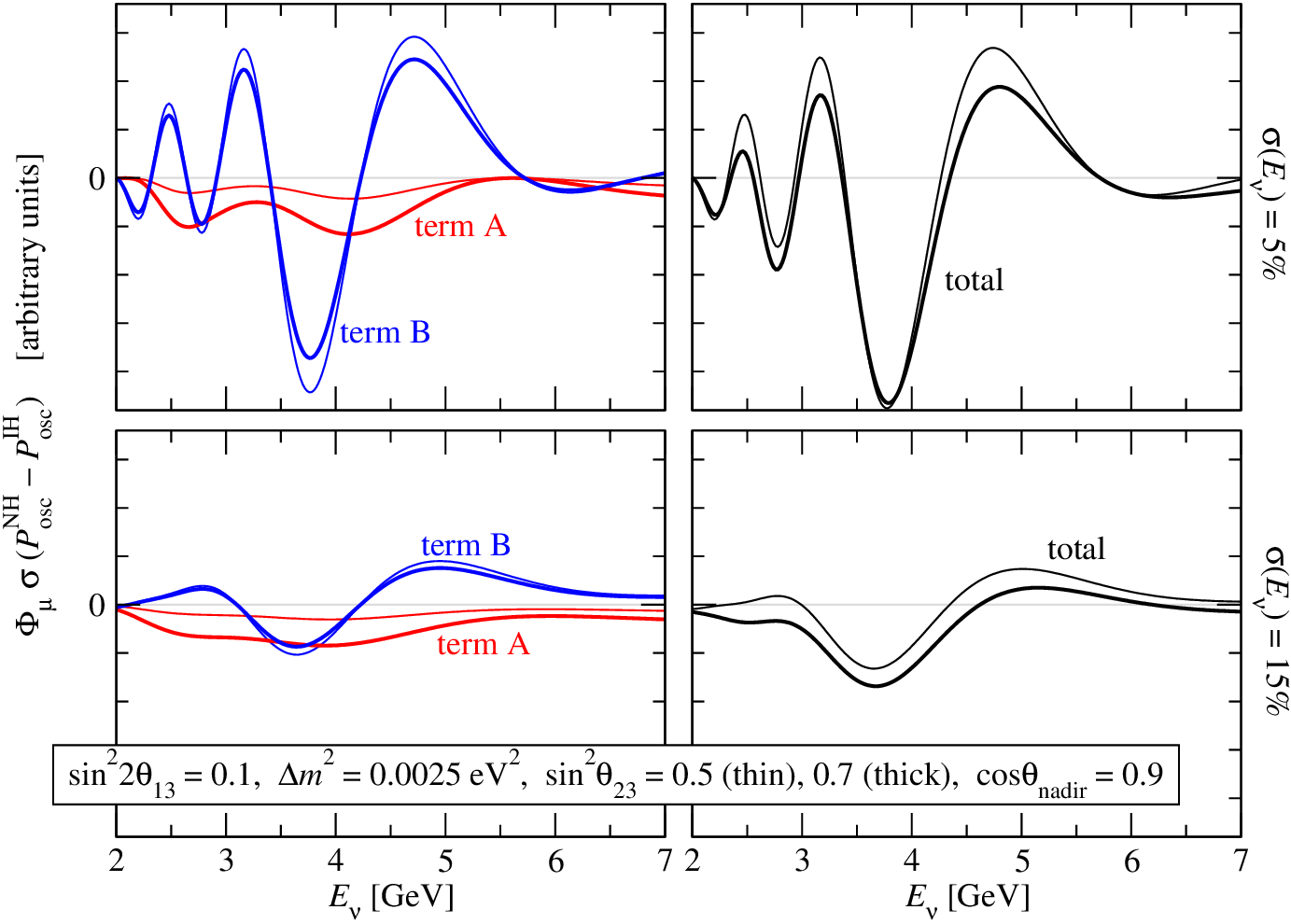}
  \mycaption{The difference between the $\mu$-like event energy
  spectra corresponding to $\Delta m^2 > 0$ (NH) and $\Delta m^2 < 0$
  (IH), $\Delta S^\nu_\mu$, defined in Eq.~(\ref{eq:diff-spect}). In
  the left panels terms A and B are displayed separately, in the right
  panels the total effect is shown.  In the upper (lower) panels an
  energy resolution of $5\%$ ($15\%$) is taken into account. The Nadir
  angle is fixed to $\cos\theta_n = 0.9$ (no smearing included). Thin
  (thick) curves correspond to $\sin^2\theta_{23} = 0.5 \,(0.7)$. }
\label{fig:distr-0.9}
\end{figure}

In order to determine the type of the neutrino mass spectrum one can
explore the difference in the event energy spectra as predicted by the NH
and IH. To illustrate the effect for $\mu$-like events, let us consider
\begin{equation}\label{eq:diff-spect}
\Delta S^p_\mu \propto
\underbrace{%
\Phi_\mu^p \sigma^p \,
\sin^2\theta_{23}
\left(\frac{1}{r^p} - \sin^2\theta_{23} \right) \Delta P^p_{2\nu}
}_{\mbox{term A}}
\:+\:
\underbrace{%
\Phi_\mu^p \sigma^p \,
\frac{1}{2}\sin^22\theta_{23} \Delta \mathrm{Re} \, (A'^p_{33})
}_{\mbox{term B}} \,,
\end{equation}
with $\Delta X \equiv X(\mathrm{NH}) - X(\mathrm{IH})$. In
Fig.~\ref{fig:distr-0.9} we show $\Delta S^\nu_\mu$, where in the left
panels we display the two terms in Eq.~(\ref{eq:diff-spect})
separately, whereas the right panels show the total effect, \ie, the
sum of ``term A'' and ``term B''. In this plot we take into account
also the effect of the finite energy resolution, \ie, we average
$\Delta S^\nu_\mu$ according to Eq.~(\ref{eq:resolutions}) assuming
$\sigma_E = 5\%\, (15\%)$ for the upper (lower) panels. Note however,
that we do not average over the Nadir angle, and we fix $\cos\theta_n
= 0.9$.

From the upper left panel we observe that term~B provides a large
signal with a fast oscillatory behaviour in neutrino energy. In
contrast, term~A is much less sensitive to the neutrino energy, in
particular its sign does not change. This implies that the averaging
implied by worse energy resolutions mainly affects term~B, whereas
term~A is rather insensitive to it. This effect can be observed in the lower
left panel, where the signal from term~B is significantly reduced with
respect to the upper panel, but the size of term~A is similar. As
visible in the right panels, the total signal is also strongly
affected by the energy averaging, since for good resolutions term~B
dominates. 
We find a qualitatively similar behaviour also for other values of the
Nadir angle, as well as for the distribution in $\cos\theta_n$ for
fixed energy, and for antineutrinos.

Furthermore, we illustrate in Fig.~\ref{fig:distr-0.9} the dependence
of the hierarchy effect on $\theta_{23}$. According to
Eq.~(\ref{eq:diff-spect}), for term~A this dependence is controlled by
the factor $\sin^2\theta_{23}(1/r^p - \sin^2\theta_{23})$, where in
the relevant range of energies $r^p \simeq 2.6 - 4.5$. Hence, term~A
increases with $\sin^2\theta_{23}$, very similar to effects in
$e$-like events (see, \eg,~\cite{JBSP203} for a detailed
discussion). In contrast, term~B depends on $\sin^22\theta_{23}$, and
therefore it has a maximum for $\theta_{23} = \pi/4$. This behaviour of
terms~A and B is visible in the left panels of
Fig.~\ref{fig:distr-0.9}.
The dependence of the total effect on $\theta_{23}$ emerges from a
non-trivial interplay of the two terms, where the effect of averaging
plays a crucial role to control the relative size of them. We will
return to this discussion in Sec.~\ref{sec:results}, where we present
the dependence of the hierarchy sensitivity as a function of
$\sin^2\theta_{23}$.

\section{Energy and Angular Resolution, Number of Bins, Charge~ID, and
Systematics} 
\indent

In this Section we discuss in some detail the impact of
``experimental'' parameters such as energy and angular resolution, the
number of bins used in the analysis, charge identification, and
systematical errors on the ability to identify the neutrino mass
hierarchy. Since our aim in this section is to identify the impact of
these parameters on the sensitivity we perform a simplified analysis.
For most purposes we take into account only statistical errors,
\ie\ we fix all the pull parameters $\boldsymbol{\xi}$ in
Eq.~(\ref{eq:chisq}) to zero. Furthermore, we compare NH and IH for
{\it fixed} oscillation parameters. More precisely, we calculate data
for the NH and the true values $\boldsymbol{x}^\mathrm{true} =
\boldsymbol{x}^\mathrm{NH} = (\theta_{13}, \theta_{23}, \Delta m^2)$
with $\Delta m^2 > 0$. Then the $\chi^2$ is calculated for the {\it
same} oscillation parameters but for IH:
\begin{equation}\label{eq:chisq-fixed}
\Delta\chi^2(\mathrm{NH;IH}) \equiv 
\chi^2(\boldsymbol{x}^\mathrm{NH}; \boldsymbol{x}^\mathrm{IH})
\end{equation}
with $\boldsymbol{x}^\mathrm{IH} = (\theta_{13}, \theta_{23}, -\Delta
m^2)$.  The total event number is normalized to 100 according to
Eq.~(\ref{eq:normalization}). It follows from Eq.~(\ref{eq:chisq})
that for statistical errors only (\ie, $\boldsymbol{\xi}=0$), the
$\chi^2$ is proportional to the total number of events, and hence, our
results can be easily scaled for any event number. The scaling in the
presence of systematical errors is discussed at the end of this section.
 
We are aware of the fact that for atmospheric neutrino experiments
charge identification is very difficult for electrons. Therefore, most
likely only data consisting of $\mu$-like events will be
available. Nevertheless, for comparison we show our results also for
an $e$-like event sample of the same size and with the same
characteristics as the $\mu$-like data. This will allow us to
highlight the advantages or disadvantages of using the $\mu$-like
events, and to discuss the different requirements on the experiments.

\begin{figure}[!t]
\centering 
\includegraphics[width=0.9\textwidth]{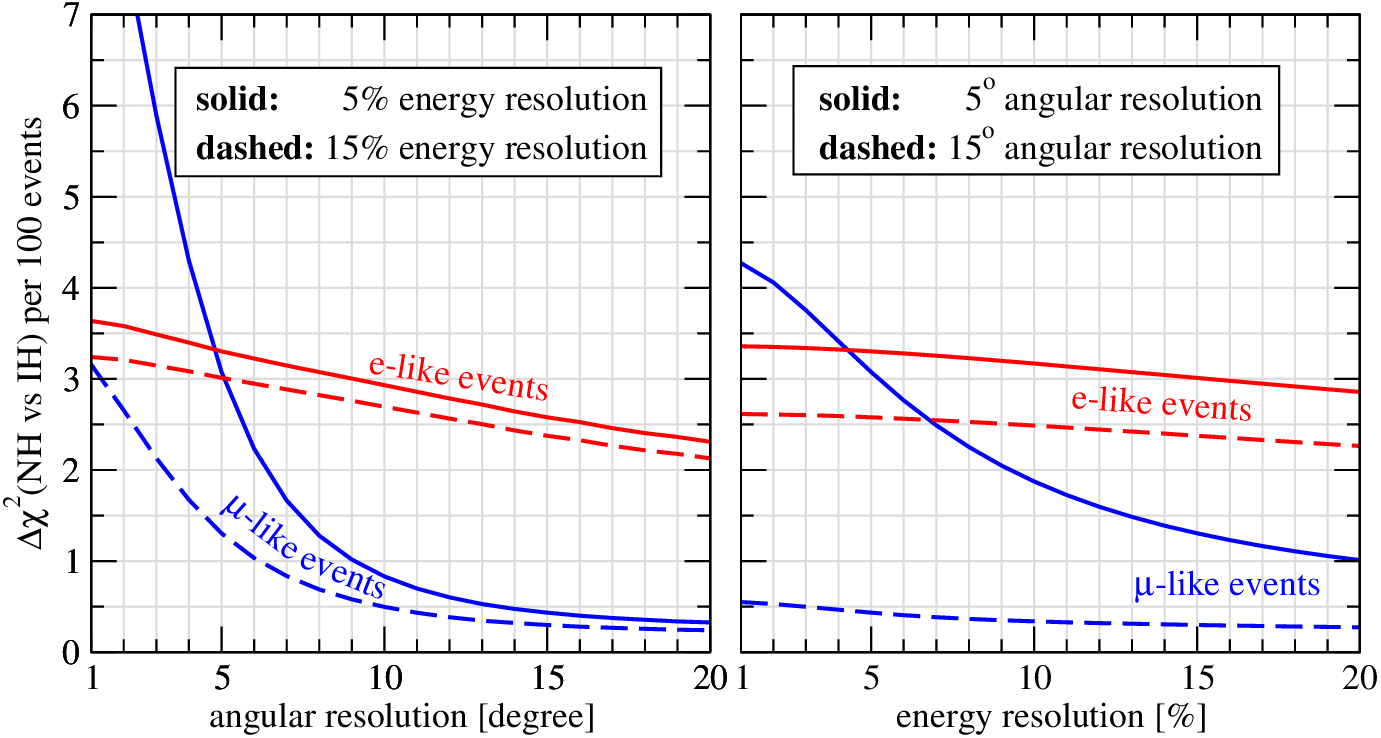}
  \mycaption{$\Delta\chi^2$ between NH and IH per 100 events as
   defined in Eq.~(\ref{eq:chisq-fixed}) as a function of the angular
   resolution (left) and energy resolution (right). The oscillation
   parameters are fixed to $\sin^22\theta_{13} = 0.1$,
   $\sin^2\theta_{23} = 0.5$, $|\Delta m^2|=2.4\times 10^{-3}$~eV$^2$,
   and we use $20\times 20$ bins in the intervals 2~GeV~$\le E_\nu \le
   10$~GeV and $0.1 \le \cos\theta_n\le 1$, statistical errors only,
   and 100\% charge identification.}
\label{fig:resolutions}
\end{figure}

Let us first explore the impact of energy and angular resolutions. In
Fig.~\ref{fig:resolutions} we show the $\Delta\chi^2$ between NH and
IH as defined in Eq.~(\ref{eq:chisq-fixed}) as a function of the
energy resolution and the angular resolution. In agreement
with~\cite{Indumathi:2004kd} we observe that for $\mu$-like events
these resolutions have a dramatic impact on the hierarchy sensitivity.
This can be understood from comparing upper and lower panels of
Fig.~\ref{fig:distr-0.9}. Due to the fast oscillations of the signal,
averaging has a rather large effect and leads to a very different
final signal. The left panel of Fig.~\ref{fig:resolutions} shows that
especially for very good angular resolutions $\sigma_\mathrm{dir}
\lesssim 5^\circ$ and energy resolutions $\sigma_E\simeq 5\%$ a very
high sensitivity to the mass hierarchy can be obtained. Even with only
100 events $\Delta \chi^2$-values corresponding to $2-3\sigma$ can be
achieved, and the sensitivity of $\mu$-like events becomes even better
than the one from $e$-like events.

In contrast, one can see from Fig.~\ref{fig:resolutions} that the
sensitivity from $e$-like events is only slightly affected by angular
and energy resolutions. The reason for this is that for $e$-like
events the effect does not show such a pronounced oscillatory shape as
in the case of $\mu$-like events, but is more similar to the
``term~A'' shown in Fig.~\ref{fig:distr-0.9}. (See~\cite{JBSP203} for
a detailed discussion of the hierarchy effect in $e$-like events in
the context of a water-Cherencov detector.) This implies that
averaging has a much smaller impact.

\begin{figure}[!t]
\centering 
\includegraphics[width=0.55\textwidth]{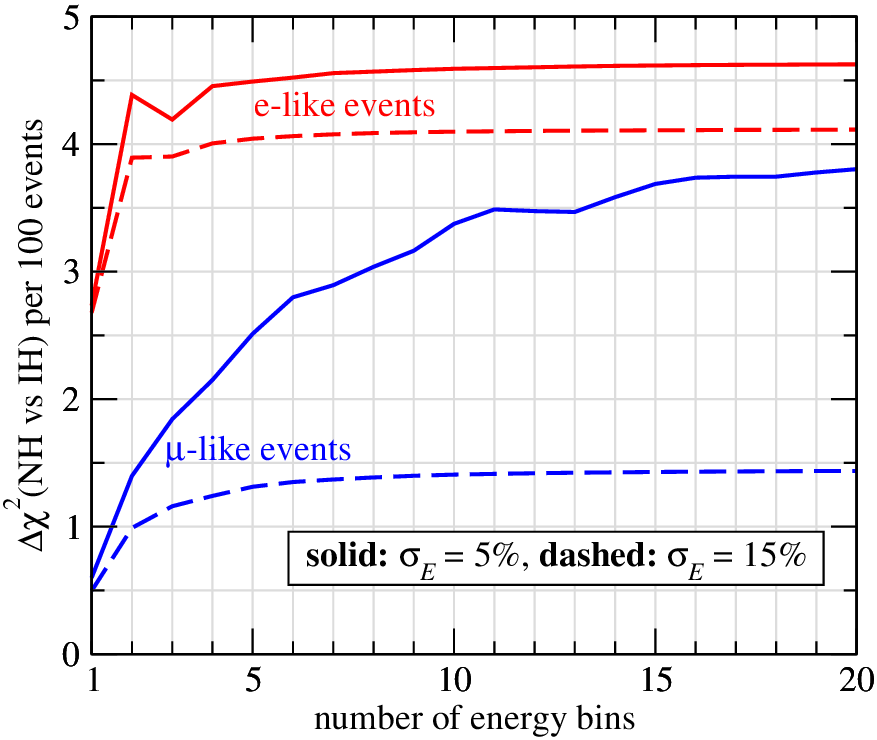}
  \mycaption{$\Delta\chi^2$ between NH and IH per 100 events as
   defined in Eq.~(\ref{eq:chisq-fixed}) as a function of the number
   of energy bins in the interval 2~GeV~$\le E_\nu \le 8$~GeV. The
   oscillation parameters are fixed to $\sin^22\theta_{13} = 0.1$,
   $\sin^2\theta_{23} = 0.5$, $|\Delta m^2|=2.4\times 10^{-3}$~eV$^2$,
   and we use 10 bins in the interval $0.3 \le \cos\theta_n\le 1$,
   $\sigma_\mathrm{dir} = 5^\circ$, statistical errors only, and 100\%
   charge identification.}
\label{fig:number-of-bins}
\end{figure}

We conclude from Fig.~\ref{fig:resolutions} that taking into account
information on the neutrino energy and direction significantly
improves the sensitivity to the neutrino mass hierarchy of $\mu$-like
events. This result is confirmed by Fig.~\ref{fig:number-of-bins},
where we show the $\Delta\chi^2(\mathrm{NH;IH})$ as a function of the
number of bins in neutrino energy $N^\mathrm{bin}_E$. The value
$N^\mathrm{bin}_E = 1$ corresponds to only a rate measurement, whereas
increasing values of $N^\mathrm{bin}_E$ imply more spectral
information. We observe from that figure that for $\mu$-like events
with good energy resolution of $\sigma_E = 5\%$ the spectral
information significantly improves the sensitivity to the mass
hierarchy.  Moving from just a rate measurement to 20 bins in $E_\nu$
increases $\Delta\chi^2$ roughly by a factor of 6. For an energy
resolution of 15\% the optimal $\Delta\chi^2$ is already reached for
approximately 8 bins, since for $N^\mathrm{bin}_E \gtrsim 8$ the
energy resolution becomes larger than the bin size, and hence finer
binning cannot increase the information. However, also in this case
appropriate binning increases $\Delta\chi^2$ approximately by a factor
of 3 compared to a pure rate measurement.
Furthermore, one finds from Fig.~\ref{fig:number-of-bins} that $e$-like
events are much less sensitive to the number of energy bins, and
already for $N^\mathrm{bin}_E \simeq 2$ the optimal $\Delta\chi^2$ is
obtained. This is in agreement with our previous observations in
relation with Fig.~\ref{fig:resolutions}, that $e$-like events are
much less affected by averaging over energy and direction.

\begin{figure}[!t]
\centering 
\includegraphics[width=0.55\textwidth]{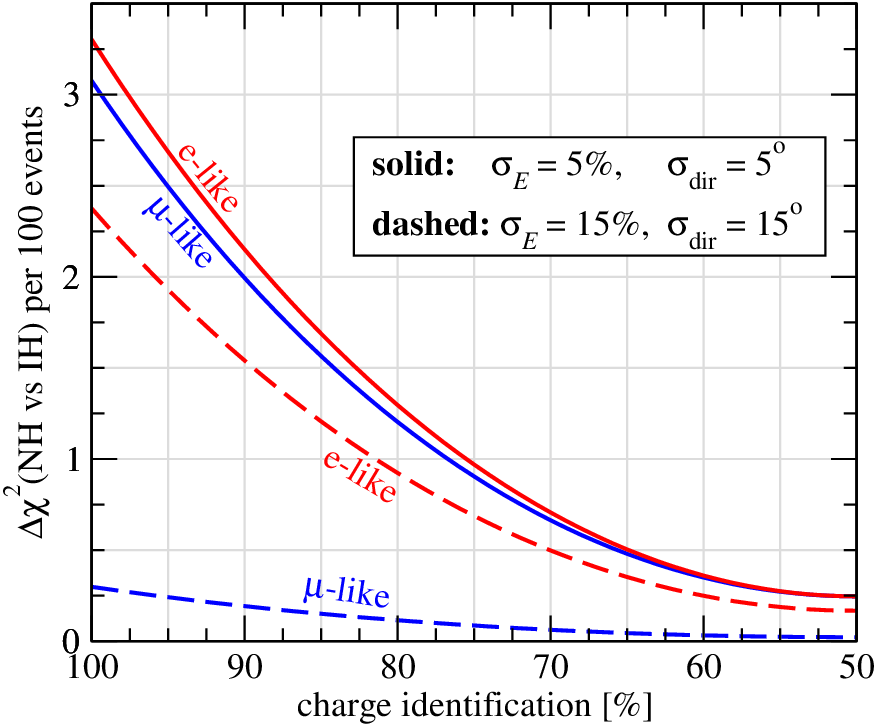}
  \mycaption{$\Delta\chi^2$ between NH and IH per 100 events as
   defined in Eq.~(\ref{eq:chisq-fixed}) as a function of the charge
   identification in percent, $100\% \times f_\mathrm{ID}$, see
   Eq.~(\ref{eq:observables}). The oscillation parameters are fixed to
   $\sin^22\theta_{13} = 0.1$, $\sin^2\theta_{23} = 0.5$, $|\Delta
   m^2|=2.4\times 10^{-3}$~eV$^2$, and we use $20\times 20$ bins in
   the intervals 2~GeV~$\le E_\nu \le 10$~GeV and $0.1 \le
   \cos\theta_n\le 1$ and statistical errors only.}
\label{fig:ch_id}
\end{figure}

In Fig.~\ref{fig:ch_id} we show the sensitivity to the mass hierarchy
as a function of the charge identification. As expected a bad charge
ID has a significant impact on the hierarchy sensitivity since,
depending on the hierarchy, the resonant matter effect occurs either
for neutrinos or antineutrinos, and hence, distinguishing neutrino
from antineutrino events is crucial. A charge ID of 90\% leads
approximately to a reduction of the $\Delta\chi^2$ of 30\% compared to
perfect charge ID. For no information on the charge ($f_\mathrm{ID} =
0.5$) the $\Delta\chi^2$ is reduced roughly by one order of magnitude.
This is one reason why in the case of a water Cherencov detector with no
possibility to distinguish neutrino from antineutrino events very
large exposures of the order of several Mt~yrs are necessary to
explore the mass hierarchy effect~\cite{Huber:2005ep}.

\begin{figure}[!t]
\centering 
\includegraphics[width=0.95\textwidth]{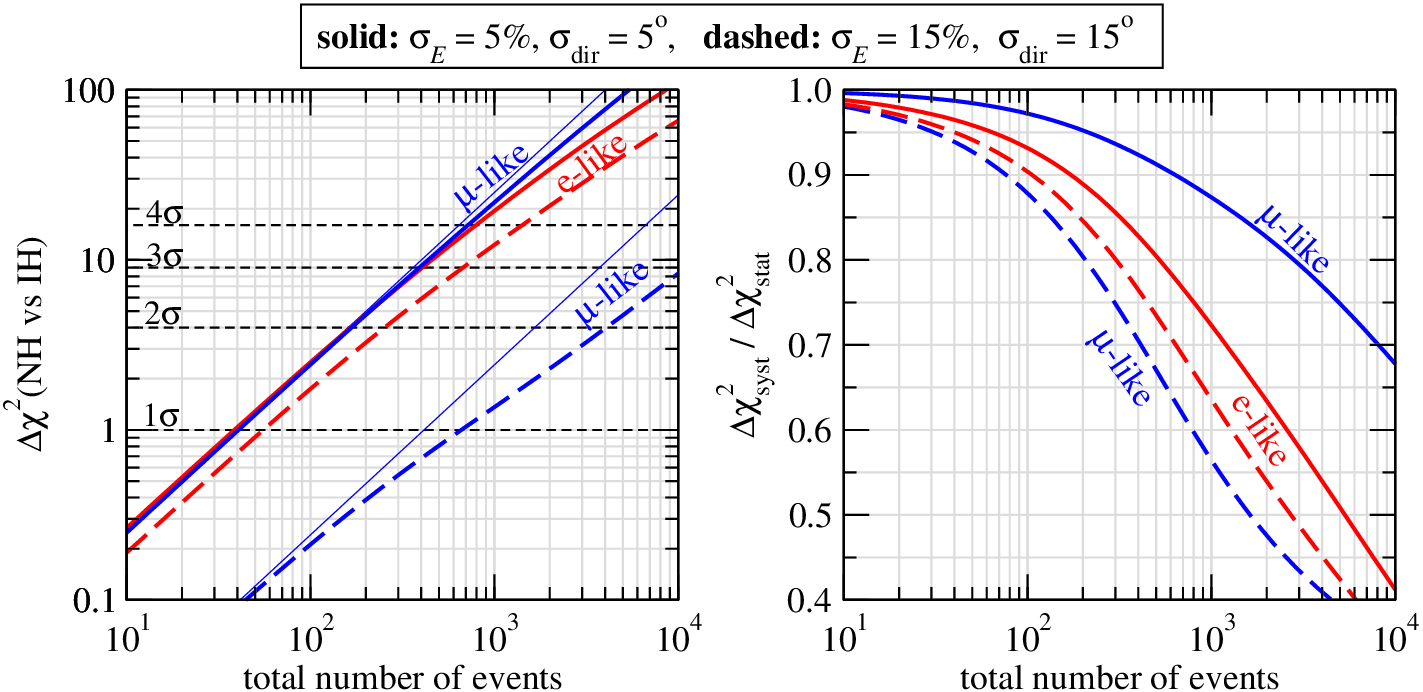}
  \mycaption{We show as a function of the total number of events the
   $\Delta\chi^2$ between NH and IH as defined in
   Eq.~(\ref{eq:chisq-fixed}) including systematical uncertainties
   (left panel), and the ratio between the
   $\Delta\chi^2(\mathrm{NH;IH})$ including systematical uncertainties
   and with statistical errors only (right panel). The values for the
   systematical uncertainties are given in Tab.~\ref{tab:systematics}.
   The thin solid lines in the left panel correspond to statistical
   errors only (shown only for $\mu$-like events). The oscillation
   parameters are fixed to $\sin^22\theta_{13} = 0.1$,
   $\sin^2\theta_{23} = 0.5$, $|\Delta m^2|=2.4\times 10^{-3}$~eV$^2$,
   and we use $20\times 20$ bins in the intervals 2~GeV~$\le E_\nu \le
   10$~GeV and $0.1 \le \cos\theta_n\le 1$, and a charge identification
   of 95\%.}
\label{fig:scaling-syst}
\end{figure}

Finally we investigate the impact of systematical errors and the
scaling of the $\chi^2$ with respect to the total number of events.
In the left panel of Fig.~\ref{fig:scaling-syst} we show the
$\Delta\chi^2$ between NH and IH as a function of the total number of
events according to Eq.~(\ref{eq:normalization}) including
systematical uncertainties. The thin solid lines show explicitly for
$\mu$-like events that in the case of no systematical errors $\chi^2$
scales linearly with the number of events. Hence, the deviation of the
curves in the left panel of Fig.~\ref{fig:scaling-syst} from straight
lines with inclination 1 is the manifestation of the systematical
errors. We conclude from this figure that---given fixed oscillation
parameters as defined in the caption---roughly 4000 events are needed
to obtain a $2\sigma$ effect for the hierarchy identification using
$\mu$-like events with resolutions of $\sigma_E = 15\%$,
$\sigma_\mathrm{dir} = 15^\circ$. In contrast, using $e$-like events
or high resolution $\mu$-like events, a $2\sigma$ sensitivity seems
possible already with $\sim 200$ events.

The impact of the systematics is shown in the right panel, where we
display the fractional decrease in $\Delta\chi^2$ by switching on
systematical uncertainties. One can read off from that plot that for
$N^\mathrm{tot} = 100$ the impact is rather small.  For the high
resolution $\mu$-like sample with $\sigma_E=5\%$ and
$\sigma_\mathrm{dir} = 5^\circ$ the $\Delta\chi^2$ is reduced only by
3\% due to the systematics; in the other shown cases the decrease in
$\Delta\chi^2$ is $\lesssim 12\%$. In this regime the $\chi^2$ is
dominated by statistical errors. However, the impact of systematics
clearly increases for larger event numbers, leading to a reduction of
$\Delta\chi^2$ up to a factor 2 for $\sim 10^4$ events.
In Fig.~\ref{fig:scaling-syst} all the systematical uncertainties
listed in Tab.~\ref{tab:systematics} are included. We have verified
that the uncertainty in the overall normalization has practically no impact,
despite the rather large value of 20\%. The 
most relevant systematics are the ones which affect the relative
number of neutrino and antineutrino events. In particular, the error
on the $\nu/\bar\nu$ ratio and the error on the charge~ID parameter
$f_\mathrm{ID}$ are important. Moreover, also the errors on the
$\cos\theta_n$ and $E_\nu$ shapes are relevant, since we take them to be
uncorrelated between neutrino and antineutrino events. In the case of
$e$-like events also the uncertainty on the $\nu_e/\nu_\mu$ flux ratio
contributes noticeable, whereas this error has practically no impact
for $\mu$-like events.

Let us comment on the relatively small effect of systematics in case
of the high resolution $\mu$-like data, visible in the right panel of
Fig.~\ref{fig:scaling-syst}. This follows from the fact that the
discrimination between NH and IH is based on a very characteristic
signal (compare Fig.~\ref{fig:distr-0.9}), consisting of pronounced
structures in the $E_\nu$ and $\cos\theta_n$ distributions, which
cannot be easily mimicked by the systematic effects.  If these
structures are washed out to some degree by the averaging implied by
worse resolutions, the impact of the systematics is increased, since
the effect of changing the hierarchy can be reduced by adjusting the
initial fluxes. The same argument applies also in the case of $e$-like
events.


\section{Results from the General Fit}
\label{sec:results}
\indent 

Before we are going to present the results of a full fit
including all parameters, we define in Tab.~\ref{tab:setups} three
benchmark setups which we will use in the following. We give in the
table the experimental characteristics used in the simulation and the
$\chi^2$ analysis. All our results in the following are normalized to
200 events for the ``true'' parameters values. A rough scaling of the
results can be performed by using Fig.~\ref{fig:scaling-syst}. The
difference between the two $\mu$-like event samples \Smuhigh\ and
\Smu\ is given by the adopted values for energy and angular
resolutions. Setup \Smuhigh\ corresponds to a ``high resolution''
sample with $\sigma_E = 5\%$ and $\sigma_\mathrm{dir} =
5^\circ$. These are rather optimistic values which might be
achievable only for a very limited number of events by applying
appropriate cuts on the muon and on the hadronic event. We discuss
setup \Smuhigh\ to illustrate the potential of an ``optimal''
experiment. The values $\sigma_E = 15\%$ and $\sigma_\mathrm{dir} =
15^\circ$ adopted for setup \Smu\ are more realistic and correspond
roughly to the numbers given in~\cite{Taba02}
referring to the MONOLITH detector.
Let us add that in a realistic experiment the $\mu$-like data will
consist of events of very diverse quality.  The accuracy with which
neutrino energy and direction can be reconstructed depends strongly on
the observed muon energy and on the details of the hadronic
event. Therefore, our Setups~A and B correspond to ideal cases of well
defined energy and direction resolutions, which nevertheless serve as
representative examples to illustrate the sensitivity.
With setup \Se\ we show also results for $e$-like data. Being aware of
the fact that charge identification is difficult for electrons we
adopt a value of 80\%. Although such a number might still be
over-optimistic, we include $e$-like data in our discussion since
already with relatively small numbers of events a good sensitivity to
the mass hierarchy can be obtained. Moreover, as we have demonstrated
in the previous section, for $e$-like events the accuracies of
neutrino energy and direction reconstruction are much less critical
than in the case of $\mu$-like data (see Fig.~\ref{fig:resolutions}).

\begin{table}[t]
\centering
\begin{tabular}{|l|ccc|}
\hline\hline
setup label & \hspace*{3ex}\Smuhigh\hspace*{3ex}
            & \hspace*{3ex}\Smu\hspace*{3ex}
            & \hspace*{3ex}\Se\hspace*{3ex}\\
\hline
event type & $\mu$-like & $\mu$-like & $e$-like \\
total number of events for $\boldsymbol{x}^\mathrm{true}$ & 
            200 & 200 & 200 \\                             
energy range      & \multicolumn{3}{|c|}{2 GeV $\le E_\nu \le 10$ GeV} \\
Nadir angle range & \multicolumn{3}{|c|}{$ 0.1 \le \cos\theta_n \le 1$} \\
number of bins in $E_\nu\times\cos\theta_n$ 
            & $20\times 20 $& $20\times 20 $& $20\times 20 $ \\
charge identification & 95\% & 95\% & 80\% \\
systematical errors & 
   \multicolumn{3}{|c|}{included according Tab.~\ref{tab:systematics}} \\
energy  resolution $\sigma_E$ & 5\% & 15\% & 15\% \\
angular resolution $\sigma_\mathrm{dir}$ 
                    & $5^\circ$ & $15^\circ$ & $15^\circ$\\ 
\hline\hline
\end{tabular}
\mycaption{Definition of example setups.
\label{tab:setups}}
\end{table} 

In the following we discuss the sensitivity of these three example
data sets to the mass hierarchy as a function of the assumed true
parameter values of $\theta_{13}$ and $\theta_{23}$. We present results
adopting three different treatments of the oscillation parameters.
First we fit the ``data'' without any additional information on the
oscillation parameters, \ie, we minimize
$\chi^2(\boldsymbol{x}^\mathrm{true}; \boldsymbol{x})$ of
Eq.~(\ref{eq:chisq}) with respect to $\boldsymbol{x}$, where the sign
of $\Delta m^2$ has opposite values in $\boldsymbol{x}^\mathrm{true}$ and
$\boldsymbol{x}$. In this case all the parameters have to be determined
by the experiment itself. 
Second, we assume that some external information on the oscillation
parameters is available when minimizing with respect to
$\boldsymbol{x}$. This information is included by adding a prior
function $\chi^2_\mathrm{prior}$ to the $\chi^2$ of
Eq.~(\ref{eq:chisq}), with
\begin{eqnarray}
\chi^2_\mathrm{prior} (\boldsymbol{x}^\mathrm{true}; \boldsymbol{x}) 
&=&
\left( \frac{|\Delta m^2| - |\Delta m^2|^\mathrm{true}}
{\sigma(\Delta m^2)} \right)^2 \nonumber\\
&+&
\left( \frac{\sin^22\theta_{23} - \sin^22\theta_{23}^\mathrm{true}}
{\sigma(\sin^22\theta_{23})} \right)^2 
+
\left( \frac{\sin^22\theta_{13} - \sin^22\theta_{13}^\mathrm{true}}
{\sigma(\sin^22\theta_{13})} \right)^2 \,, \label{eq:prior}
\end{eqnarray}
and we are using the following representative errors at $1\sigma$:
\begin{equation}\label{eq:errors}
\sigma(\Delta m^2) = 0.1 \, |\Delta m^2|^\mathrm{true} \,,\quad
\sigma(\sin^22\theta_{23}) = 0.1 \, \sin^22\theta_{23}^\mathrm{true} \,,\quad
\sigma(\sin^22\theta_{13}) = 0.02 \,.
\end{equation}
These accuracies on $|\Delta m^2|$ and $\sin^22\theta_{23}$ can be
obtained at the MINOS long-baseline and Super-K atmospheric neutrino
experiments, whereas the indicated value for
$\sigma(\sin^22\theta_{13})$ can be reached in the Double-Chooz
reactor experiment~\cite{TMU04}. Note that in the way the information
on $\theta_{23}$ and $\theta_{13}$ is included in Eq.~(\ref{eq:prior})
we take into account that $\nu_e$ ($\nu_\mu$) disappearance
experiments meassure actually $\sin^22\theta_{13}$
($\sin^22\theta_{23}$). In particular, this implies that we include no
prior information on the octant of $\theta_{23}$.
And third, we also show our results by fixing the oscillation
parameters to the true values, \ie, without minimizing the $\chi^2$
with respect to $\boldsymbol{x}$. This corresponds to the ideal case
of infinite precision on $|\Delta m^2|$, $\theta_{23}$ and
$\theta_{13}$.

\begin{figure}[!t]
\centering 
\includegraphics[width=0.95\textwidth]{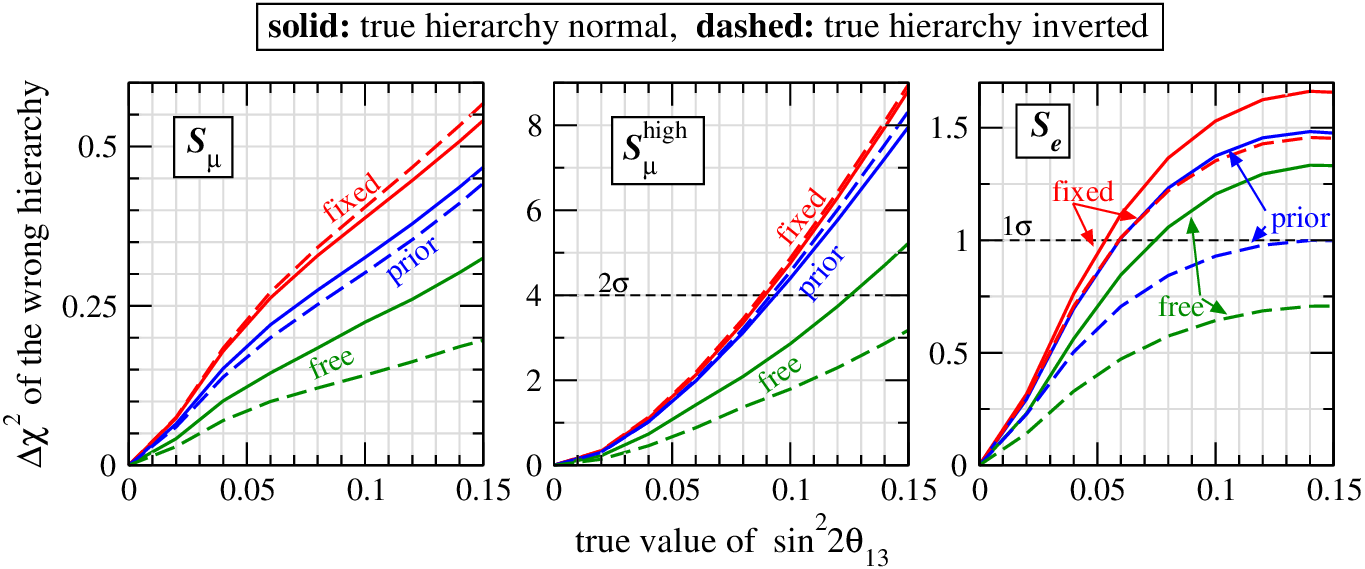}
  \mycaption{$\Delta\chi^2$ of the wrong hierarchy as a function of
  the true value of $\sin^22\theta_{13}$ for the three data samples
  defined in Tab.~\ref{tab:setups}, each consisting of 200 events:
  $\mu$-like data \Smu\ (left), $\mu$-like data with high energy and
  angular resolution \Smuhigh\ (middle), and $e$-like data \Se\
  (right). The assumed true parameter values are $|\Delta m^2| =
  2.4\times 10^{-3}$~eV$^2$, $\sin^2\theta_{23} = 0.5$, and for the
  solid (dashed) lines the true hierarchy is normal (inverted). For
  the curves labeled ``fixed'' the oscillation parameters are fixed to
  the true values, for the curves labeled ``prior'' external
  information is included according to Eq.~(\ref{eq:errors}), and for
  the curves labeled ``free'' no additional information is assumed.}
\label{fig:true_13}
\end{figure}

\begin{figure}[!t]
\centering 
\includegraphics[width=0.5\textwidth]{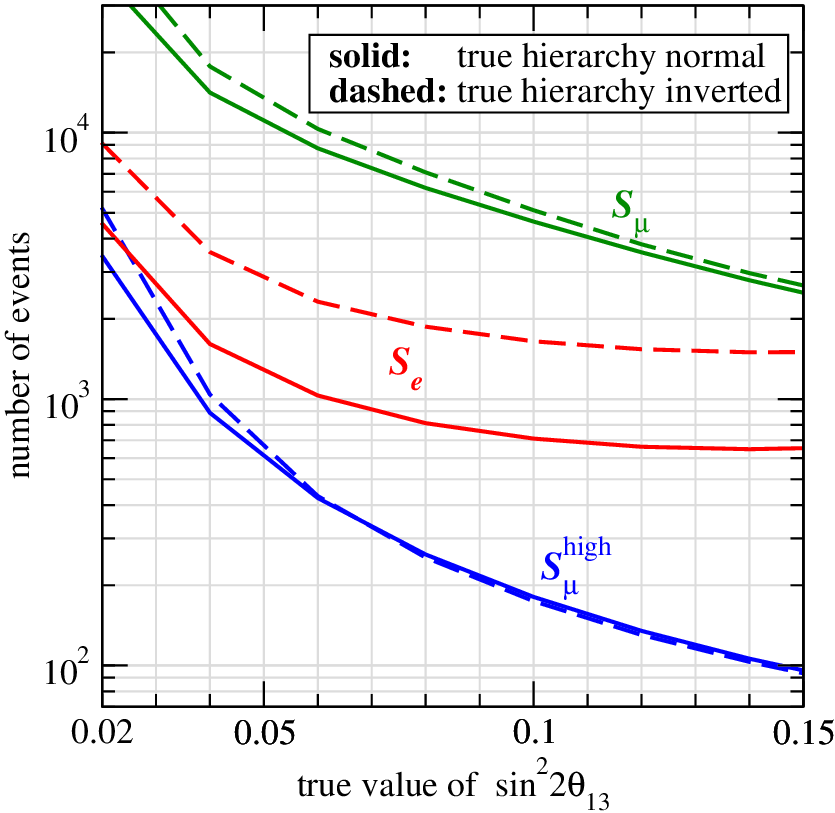}
  \mycaption{The number of events needed to exclude the wrong
  hierarchy at $2\sigma$ ($\Delta\chi^2 = 4$) as a function of the
  true value of $\sin^22\theta_{13}$ for the three data samples
  defined in Tab.~\ref{tab:setups}: $\mu$-like data \Smu, $\mu$-like
  data with high energy and angular resolution \Smuhigh, and $e$-like
  data \Se. The assumed true parameter values are $|\Delta m^2| =
  2.4\times 10^{-3}$~eV$^2$, $\sin^2\theta_{23} = 0.5$, and for the
  solid (dashed) lines the true hierarchy is normal (inverted). For
  the oscillation parameters external information is included
  according to Eq.~(\ref{eq:errors}).}
\label{fig:true_13-events}
\end{figure}

\begin{figure}[!t]
\centering 
\includegraphics[width=0.95\textwidth]{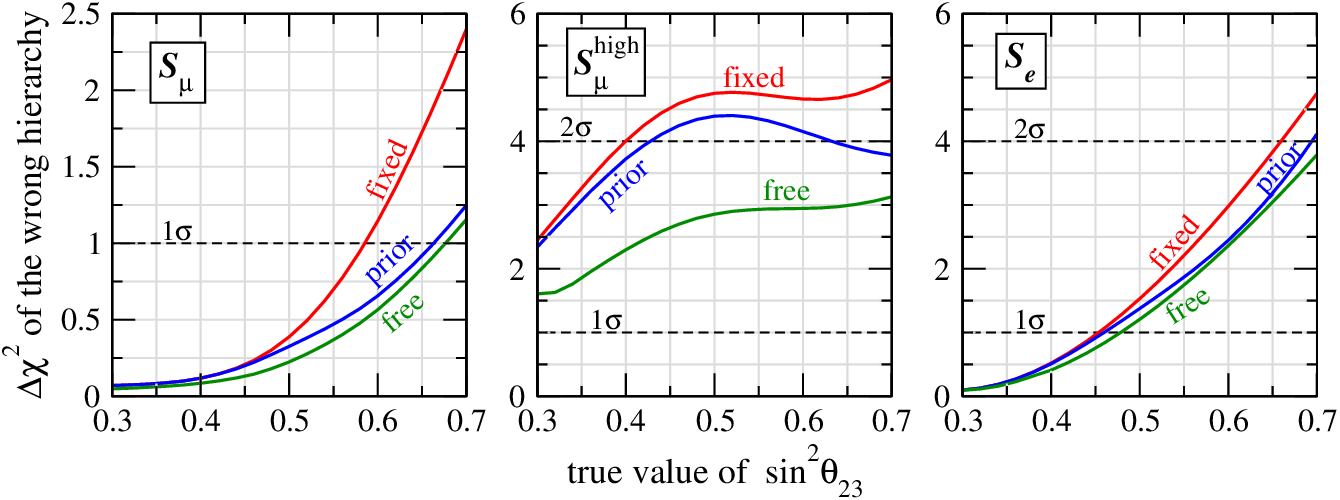}
  \mycaption{$\Delta\chi^2$ of the wrong hierarchy as a function of
  the true value of $\sin^2\theta_{23}$ for the three data samples
  defined in Tab.~\ref{tab:setups}, each consisting of 200 events:
  $\mu$-like data \Smu\ (left), $\mu$-like data with high energy and
  angular resolution \Smuhigh\ (middle), and $e$-like data \Se\
  (right). The assumed true parameter values are $\sin^22\theta_{13} =
  0.1$ and $\Delta m^2 = 2.4\times 10^{-3}$~eV$^2$ (the true
  hierarchy is normal). For the curves labeled ``fixed'' the
  oscillation parameters are fixed to the true values, for the curves
  labeled ``prior'' external information is included according to
  Eq.~(\ref{eq:errors}), and for the curves labeled ``free'' no
  additional information is assumed.}
\label{fig:true_23}
\end{figure}

The results of such analyzes are presented in Figs.~\ref{fig:true_13},
\ref{fig:true_13-events} and \ref{fig:true_23}. In
Fig.~\ref{fig:true_13} we show the $\Delta\chi^2$ of the wrong
hierarchy as a function of the true value of
$\sin^22\theta_{13}$. Clearly, a sizable value of $\theta_{13}$ is
required since the effect disappears for $\theta_{13}\to
0$. Furthermore, one observes from this figure that external
information on the oscillation parameters improves the mass hierarchy
sensitivity: Without any external constraints (curves labeled
``free'') the $\Delta\chi^2$ is significantly reduced compared to
fixed parameters. It is interesting to note that for the high
resolution sample \Smuhigh\ the modest precision implied by
Eq.~(\ref{eq:errors}) leads already to a sensitivity very close to the
one for perfectly known parameters (compare the curves labeled
``fixed'' and ``prior''). Also shown in Fig.~\ref{fig:true_13} is the
dependence of the hierarchy sensitivity on the type of the {\it true}
hierarchy (compare solid and dashed curves). One observes that if
external information on the oscillation parameters is available for
$\mu$-like data, the sensitivity is very similar for a true NH and
IH. In contrast, for $e$-like data the sensitivity is always better
for a true NH. 

In Fig.~\ref{fig:true_13-events} we answer the question of how many
events are needed to exclude the wrong hierarchy at $2\sigma$.
Considering true values $\sin^22\theta_{13} \sim 0.1$ we find that for
$\mu$-like data \Smu\ event numbers of order few$\times 10^3$ are
needed, whereas in case of high resolution data \Smuhigh\ already
$\sim 200$ events are sufficient. For $e$-like data
$\mathcal{O}(10^3)$ events are required.
For comparison, from
Refs.~\cite{Taba02,Indumathi:2004kd} we roughly
estimate 4 events per kton year in a MONOLITH/INO type
detector. Hence, with the proposed detector mass of 30~kton and a 5
year exposure $\sim 600$ events are obtained. We conclude that for the
15\% energy and $15^\circ$ angular resolutions adopted for \Smu\ only
a very poor sensitivity to the mass hierarchy will be
reachable. Therefore, improving the resolutions beyond these values and/or
significantly larger exposures seem to be necessary.

Finally we discuss the sensitivity to the mass hierarchy as a function
of the true value of $\sin^2\theta_{23}$. So far we always have
assumed maximal mixing $\sin^2\theta_{23} = 0.5$. As shown in
Fig.~\ref{fig:true_23}, the ability to identify the true hierarchy
depends quite significantly on the true value of $\theta_{23}$, and in
general the $\Delta\chi^2$ of the wrong hierarchy increases for
increasing $\sin^2\theta_{23}$. For example, for the data sample \Smu\
and fixed oscillation parameters, $\Delta\chi^2$ increases by a factor
3 if $\sin^2\theta_{23}$ is moved from 0.5 to 0.6. However, we note
from the left panel of Fig.~\ref{fig:true_23} that for \Smu\ and
$\sin^2\theta_{23} > 0.5$ the uncertainty on the oscillation
parameters has a strong impact on the final sensitivity. The reason is
a degeneracy between the octants of $\theta_{23}$: For a true
$\theta_{23} > \pi/4$ the data can be fitted better with the wrong
hierarchy for $\theta'_{23} \approx \pi/2 - \theta_{23}$. Note that
our prior function Eq.~(\ref{eq:prior}) is insensitive to the octant
of $\theta_{23}$. We have checked that if in addition to the
accuracies given in Eq.~(\ref{eq:errors}) also the octant of
$\theta_{23}$ were known, results close to the case of fixed
oscillation parameters could be obtained \footnote{The possibility of
a precision measurement of $\theta_{23}$ in an atmospheric neutrino
experiment with an iron magnetized detector has been studied recently
in~\cite{Choubey:2005zy}.}

For the high resolution data \Smuhigh\ shown in the middle panel of
Fig.~\ref{fig:true_23} we find that the maximal sensitivity is reached
around maximal mixing $\theta_{23} \approx \pi/4$.  As illustrated in
Fig.~\ref{fig:distr-0.9}, in this case the ``term B'' contributes
significantly to the sensitivity, and the behaviour shown in
Fig.~\ref{fig:true_23} follows from an interplay of ``term A'' (which
is increasing with $\sin^2\theta_{23}$) and ``term B'' (which is
proportional to $\sin^22\theta_{23}$ and hence has a maximum at
$\sin^2\theta_{23} = 0.5$).
In contrast, in the case of \Smu\ the ``term B'' is suppressed due to
the averaging implied by the worse resolutions, and hence, the
increase of the sensitivity follows from the
$\sin^2\theta_{23}$-dependence of ``term A'' shown in
Eq.~(\ref{eq:diff-spect}). Also for $e$-like events the effect is
similar to ``term A'', and according to Eq.~(\ref{eq:spectrum_e}) it is
proportional to $(r^p \sin^2\theta_{23} - 1)$. As expected, we observe
in the right panel of Fig.~\ref{fig:true_23} an increase of the
sensitivity with $\sin^2\theta_{23}$, in agreement with
Refs.~\cite{JBSP203,Huber:2005ep}.


\section{Conclusions}
\indent 

In this work we have reconsidered the possibility to determine the
type of the neutrino mass hierarchy by using atmospheric neutrinos in
a detector capable to distinguish between neutrino and antineutrino
events. For not too small values of the mixing angle $\theta_{13}$,
Earth matter effects will lead to different signals in such a
detector for a normal or inverted neutrino mass hierarchy. 
Having in mind magnetized iron calorimeters like MINOS, INO, or perhaps
ATLAS, we have performed a $\chi^2$-analysis including realistic
neutrino fluxes, detection cross sections, and various systematical
uncertainties. We discuss how the performance depends on detector
characteristics like energy and direction resolutions or charge
miss-identification, and on the systematical uncertainties related to
the atmospheric neutrino fluxes. Moreover, we show how the mass
hierarchy determination depends on the true values of $\theta_{13}$,
$\theta_{23}$, as well as on the true hierarchy. We also compare the
potential of $\mu$-like events to a (hypothetical) data sample of
$e$-like events with similar characteristics.

Focusing on the detection of muons, one of our main findings is that
the ability to reconstruct the energy and direction of the neutrino is
crucial for the mass hierarchy determination. Assuming
$\sin^22\theta_{13} \simeq 0.1$ and the optimistic values for the
neutrino energy and direction reconstruction accuracies of $5\%$ and
$5^\circ$, respectively, the mass hierarchy can be identified at the
2$\sigma$~C.L.\ already with roughly 200 events (sum of neutrino and
antineutrino events, including oscillations).  In contrast, for less
ambitious resolutions of $15\%$ and $15^\circ$, of the order of 5000
events are needed. These numbers are based on the detection of muons
with a correct charge identification of $95\%$ and an energy threshold
of 2~GeV.
For illustration we have also considered the potential of using
$e$-like event samples. Assuming energy and direction resolutions of
$15\%$ and $15^\circ$ and a correct charge identification of $80\%$ we
find that of order 1000 events are necessary for a 2$\sigma$ hierarchy
determination.

The reason for the relatively high sensitivity which can be achieved
using high resolution $\mu$-like events comes from the fact that the
difference in the signals for normal and inverted hierarchies shows a
characteristic oscillatory behaviour in the neutrino energy, as well
as in the Nadir angle. If the energy and direction reconstruction are
sufficiently precise to resolve these structures, a statistical
analysis using energy as well as directional information (ideally an
un-binned likelihood analysis) provides very good sensitivity to the
hierarchy. For worse energy and direction reconstruction, these
oscillatory structures are averaged out, which leads to a much worse
sensitivity.
Our results imply that for resolutions of $15\%$ for the neutrino
energy and $15^\circ$ for the neutrino direction, in a
MONOLITH/INO-like detector exposures of the order of a few Mton years
are required to obtain a reasonable hierarchy sensitivity. However, we
stress again that for high quality event samples with more precise
reconstruction of the neutrino energy and direction the required
exposures are significantly smaller.

\vspace*{5mm}{\bf Acknowledgments.} 
We thank P.~Huber, M.~Maltoni, S.~Palomares-Ruiz and F.~Vannucci for
useful discussions. This work was supported in part by the Italian
MIUR and INFN under the programs ``Fisica Astroparticellare''
(S.T.P.). The work of T.S.\ is supported by a ``Marie Curie
Intra-European Fellowship within the 6th European Community Framework
Program.''

\end{document}